\documentclass[journal]{IEEEtran}

\usepackage{cite}
\usepackage[numbers,sort&compress]{natbib}
\usepackage{ragged2e}
\usepackage{color}

\ifCLASSINFOpdf
   \usepackage[pdftex]{graphicx}
\else
\fi
\usepackage{amsmath}
\usepackage{amsmath,bm}
\usepackage{amsthm,amssymb}
\usepackage{mathrsfs}
\usepackage{textcomp}
\usepackage[dvipsnames]{xcolor}
\usepackage[most]{tcolorbox}
\usepackage[ruled,linesnumbered]{algorithm2e}
\allowdisplaybreaks[4]

\ifCLASSOPTIONcompsoc
  \usepackage[caption=false,font=normalsize,labelfont=sf,textfont=sf]{subfig}
\else
  \usepackage[caption=false,font=footnotesize]{subfig}
\fi

\usepackage{fixltx2e}

\usepackage{nomencl}
\usepackage{etoolbox}
\usepackage{graphicx}
\usepackage{bbding}
\usepackage{float}

\newtheorem{proposition}{Proposition}

\renewcommand\nomgroup[1]{%
  \item[\bfseries
  \ifstrequal{#1}{B}{Sets}{%
  \ifstrequal{#1}{F}{First-stage Decision Variables}{%
  \ifstrequal{#1}{G}{Second-stage Decision Variables}{%
  \ifstrequal{#1}{R}{Random Variables}{%
  \ifstrequal{#1}{C}{Constant parameters}{%
  \ifstrequal{#1}{A}{Indices}{}}}}}}%
]}
\makenomenclature

\makeatletter
\renewcommand{\maketag@@@}[1]{\hbox{\m@th\normalsize\normalfont#1}}%
\makeatother

\begin{document}
\title{Adaptive Distributionally Robust Planning for Renewable-Powered Fast Charging Stations Under
Decision-Dependent EV Diffusion Uncertainty}

\author{Yujia Li,~\IEEEmembership{Member,~IEEE,}
         Feng Qiu,~\IEEEmembership{Senior Member,~IEEE,}
        Yixuan Chen,~\IEEEmembership{Member,~IEEE,}
      Yunhe Hou,~\IEEEmembership{Senior Member,~IEEE}}
\maketitle

\begin{abstract}
When deploying fast charging stations (FCSs) to support long-distance trips of electric vehicles (EVs), there exist indirect network effects: while the gradual diffusion of EVs directly influences the timing and capacities of FCS allocation, the decisions for FCS allocations, in turn, impact the drivers' willingness to adopt EVs. 
This interplay, if neglected, can result in uncovered EVs and security issues on the grid side and even hinder the effective diffusion of EVs. 
In this paper, we explicitly incorporate this interdependence by quantifying EV adoption rates as decision-dependent uncertainties (DDUs) using decision-dependent ambiguity sets (DDASs). Then, a two-stage decision-dependent distributionally robust FCS planning (D$^3$R-FCSP) model is developed for adaptively deploying FCSs with on-site sources and expanding the coupled distribution network. 
A multi-period capacitated arc cover-path cover (MCACPC) model is incorporated to capture the EVs' recharging patterns to ensure the feasibility of FCS locations and capacities. 
To resolve the nonlinearility and nonconvexity, 
the D$^3$R-FCSP model is equivalently reformulated into a single-level mixed-integer linear programming by exploiting its strong duality and applying the McCormick envelop. Finally, case studies highlight the superior out-of-sample performances of our model in terms of security and cost-efficiency. 
Furthermore, 
the byproduct of accelerated EV adoption through an implicit positive feedback loop is highlighted.

\end{abstract}

\begin{IEEEkeywords}
Coupled transportation and power systems, decision-dependent uncertainty,   distributionally robust optimization, fast charging station, flow refueling location model.

\end{IEEEkeywords}

\IEEEpeerreviewmaketitle
\mbox{}

{\small
\printnomenclature[1in]}
\nomenclature[A]{$n$}{Index of DN node $n$}
\nomenclature[A]{$od$}{Index of OD pairs (and the path between them)}
\nomenclature[A]{$\gamma$}{Index of planning periods}
\nomenclature[A]{$i$}{Index of TN node $i$}
\nomenclature[A]{$(j,k)$}{Index of a directional arc from node $j$ to node $k$ in the TN}
\nomenclature[A]{$s$}{Index of scenarios for EV adoption rates}
\nomenclature[A]{$d$}{Index of representative days}
\nomenclature[A]{$(n,m)$}{Index of directed distribution line from DN node $n$ to $m$}
\nomenclature[B]{$\mathcal{T}$}{Set of time slots} 
\nomenclature[B]{$\mathcal{D}_\gamma$}{Set of representative days at period $\gamma$}
\nomenclature[B]{$\Gamma$}{Set of planning periods}
\nomenclature[B]{$\mathcal{N}^{\rm TN}$}{Set of all the nodes in the TN}
\nomenclature[B]{$\mathcal{L}$}{Set of distribution lines}
\nomenclature[B]{$\mathcal{K}_{od}(j,k)$}{Set of candidate nodes that can cover arc $(j,k)$ 
\nomenclature[B]{$\mathcal{Q}$}{Set of OD pairs $od$}
\nomenclature[B]{$\mathcal{A}^{\rm TN}_{od}$}{Set of directional arcs on path $od$, sorted from origin to destination and back to origin}
\nomenclature[B]{$\mathcal{N}^{\rm TN}_{od}$}{Set of ordered nodes on path $od$}
\nomenclature[B]{$\mathcal{N}^{\rm DN}$}{Set of all nodes in the DN}
\nomenclature[B]{$\mathcal{N}^{\rm D\text{-}T}_n$}{Set of TN nodes connected to DN node $n$}
\nomenclature[B]{$\mathcal{S}_\gamma$}{Set of realizations of EV adoption rates $s$}
\nomenclature[F]{$x^{\rm ch}_{i,\gamma}$}{Binary variable indicating the existence of an FCS; equals 1 if an FCS exists}
\nomenclature[F]{$\hat{x}^{\rm ch}_{i,\gamma}$}{Binary variables indicating the newly constructed FCS; equals 1 if an FCS is newly constructed}
\nomenclature[F]{$P^{\rm re}_{i,\gamma}$}{Overall installed power capacity of PVs}
\nomenclature[F]{$E^{\rm es}_{i,\gamma}$}{Overall installed energy capacity of ESSs}
\nomenclature[F]{$x^{\rm L}_{nm,\gamma}$}{Binary variable indicating the constructed conductor; equals 1 if the conductor has been expanded}
\nomenclature[F]{$\hat{x}^{\rm L}_{nm,\gamma}$}{Binary variable indicating the newly expanded conductor; equals 1 if the conductor is newly expanded}
\nomenclature[F]{$z^{\rm ch}_{i,\gamma}$}{Integer variable indicating the number of CSs}
\nomenclature[F]{$\hat{z}^{\rm ch}_{i,\gamma}$}{Integer variable indicating the newly installed number of CSs}
\nomenclature[F]{$\hat{P}^{\rm re}_{i,\gamma}, \hat{E}^{\rm es}_{i,\gamma}$}{Newly installed capacity of PVs and ESSs}
\nomenclature[F]{$\hat{P}^{\rm sub}_{i,\gamma}$}{Expanded capacity of a substation}
\nomenclature[G]{$p^{\rm line}_{nm,d,t},q^{ \rm line}_{nm,d,t}$}{Active/reactive power on DN line}
\nomenclature[G]{$fr_{od,i,d,t}$}{Fraction of EVs between OD pair $od$ charging at FCS $i$ on day $d$ in time slot $t$}
\nomenclature[G]{$\lambda_{i,d,t}$}{Number of EVs served by the FCS}
\nomenclature[G]{$\lambda_{i,d,t}^{\rm un}$}{Number of unsatisfied EVs}
\nomenclature[G]{$p^{\rm up}_{n,d,t},q^{\rm up}_{n,d,t}$}{Purchased active/reactive power from the main grid}
\nomenclature[G]{$p^{\rm re,cu}_{n,d,t}$}{Curtailed RES capacity}
\nomenclature[G]{$\zeta_{n,d,t}$}{Load shedding coefficient}
\nomenclature[G]{$e^{\rm es}_{i,d,t}$}{Residual energy of the ESSs}
\nomenclature[G]{$p^{\rm es,c}_{i,d,t},p^{\rm es,d}_{i,d,t}$}{Charging and discharging power of ESSs}
\nomenclature[G]{$p^{\rm re}_{n,d,t},q^{\rm re}_{n,d,t}$}{Active/reactive generation of PVs}
\nomenclature[G]{$u^{\rm sqr}_{n,d,t}$}{Squared voltage magnitude of DN node}
\nomenclature[C]{$\overline{Z}^{\rm ch}_{i,\gamma},\underline{Z}^{\rm ch}_{i,\gamma}$}{Maximum/minimum CS number in the FCS}
\nomenclature[C]{$\eta^{\rm ev}$}{Charging efficiency of the CS}
\nomenclature[C]{$\overline{U}^{\rm sqr}_n,\underline{U}^{\rm sqr}_n$}{Maximum/minimum squared voltage of DN node}
\nomenclature[C]{$R_{nm},X_{nm}$}{Resistance and reactance of the distribution line}
\nomenclature[C]{$Ed$}{Energy consumption of EVs (kWh/km)}
\nomenclature[C]{$\overline{P}^{\rm ch}$}{Rated charging power of CSs}
\nomenclature[C]{$\overline{P}^{\rm re}_i$}{Maximum installation capacity of on-site PVs}
\nomenclature[C]{$\overline{E}^{\rm es}_i$}{Maximum installation capacity of on-site ESSs}
\nomenclature[C]{$\overline{P}^{\rm L}/\overline{Q}^{\rm L}$}{Expansion capacity of the distribution line}
\nomenclature[C]{${P}^{\rm sub,0}_i$}{Initial power capacity of the substation}
\nomenclature[C]{$W_d$}{Weight of the representative day $d$}
\nomenclature[C]{$C^{\rm in}$}{Unit construction cost of an FCS}
\nomenclature[C]{$C^{\rm ch}$}{Unit installation cost of a CS}
\nomenclature[C]{${\iota}^{\rm c}/{\iota}^{\rm d}$}{Per unit charging/discharging rates of ESSs}
\nomenclature[C]{$\eta^{\rm c}/\eta^{\rm d}$}{Charging and discharging efficiency of ESSs}
\nomenclature[C]{$D^{\rm ev}$}{Driving range of an EV}
\nomenclature[C]{$C^{\rm un}$}{Unit penalty cost of uncovered EVs}
\nomenclature[C]{$C^{\rm cu},C^{\rm sh}$}{Unit penalty costs of PV curtailment and load shedding}
\nomenclature[C]{$C^{\rm re},C^{\rm es}$}{Unit installation costs of PV/ESS}
\nomenclature[C]{$C^{\rm L}$}{Unit expansion cost for distribution line (\$/km)}
\nomenclature[C]{$C^{\rm sub}$}{Unit expansion cost for substation capacity}
\nomenclature[C]{$C^{\rm up,p},C^{\rm up,q}$}{Unit purchase price of active/reactive energy from the upstream main grid}
\nomenclature[C]{$\overline{P}^{\rm line}_{nm}/\overline{Q}^{\rm line}_{nm}$}{Active/reactive capacity of the distribution line}
\nomenclature[R]{$\tilde{\theta}_{od,\gamma}$}{Adoption rate of EVs}
\nomenclature[R]{$\tilde{\Lambda}_{od,t}$}{Traffic flow volume of all vehicles}
\nomenclature[R]{$\tilde{\varpi}_{i,t}$}{Per unit PV generation}
\nomenclature[R]{$\tilde{P}^{\rm load}_{n,t}$}{Conventional load demand}
\section{Introduction} 

\IEEEPARstart{T}{he} global number of EVs has risen to around 16.5 million, nearly triple the number in 2018 \cite{IEA}. 
While the swift expansion of the EV market greatly advances global decarbonization, it also necessitates extensive updates to both transportation networks (TNs) and distribution networks (DNs). 
For TNs, the limited driving range of current EVs calls for strategically deployed fast charging stations (FCSs) to alleviate range anxiety \cite{ko2017locating} and enable long-distance travel. 
For DNs, 
the EV integration leads to a considerable rise in energy consumption and a 
redistribution of power flows, necessitating the selection of suitable coupling nodes and grid-side investments. 
These have motivated extensive research on FCS planning (FCSP) problems from the standpoint of
coupled transportation and distribution networks (TDNs). 

As DN expansions are often costly and physically infeasible, recent studies have highlighted the on-site distributed energy resources (DERs) in mitigating power congestion, voltage violations, and costly investments resulting from widespread EV integration.  
For instance, \cite{xie2018planning} explored a two-stage siting and sizing problem for FCSs fully powered by on-site photovoltaic (PV) panels on highways to support remote stand-alone areas. In \cite{shao2020coordinated}, on-site energy storage systems (ESSs) are integrated to accommodate the intra-day fluctuations of charging demands. 
In \cite{moradzadeh2020new}, 
a new mixed-integer linear programming (MILP) formulation for coordinating renewable energy sources (RESs) and ESSs into FCSs is proposed to maximize positive environmental impacts. 
Considering that long-distance trips often rely on highways with ample space, incorporating on-site DERs into FCSs becomes a feasible and promising alternative to costly grid expansions.



While deploying FCSs with on-site sources can greatly facilitate EV integration, the expanding EV market exhibits 
multi-scale uncertainties, thereby complicating the FCSP problem. 
Specifically, EV charging demands in the FCSP problem are jointly influenced by two stochastic factors on different timescales, i.e., short-scale recharging behaviors and the long-scale EV diffusion process. 
The stochasticity of the former is usually incorporated when modeling spatio-temporal transportation behaviors, 
through equilibrium-based \cite{tao2022adaptive}, flow-based methods \cite{zhang2017second}, \cite{zhang2017incorporating}, etc. 
In equilibrium-based methods, the impacts of FCS locations on EV drivers’ travel choices and 
the equilibrium flow distribution are considered. 
In \cite{tao2022adaptive}, a stochastic user equilibrium model is adopted to assign EV traffic flow considering travelers' random route choices. 
Flow-based methods are also extensively employed due to their practicality from a utility standpoint. 
In \cite{zhang2017second}, a capacitated flow refueling location model (FRLM) explicitly captures time-varying charging demands given EVs' driving ranges. 
The arc cover-path cover (ACPC) model \cite{abdalrahman2019pev} 
is another computationally efficient flow-based method, 
based on the idea that FCSs should make all the arcs on the path traversable for EVs.

On the other hand, quantifying the long-term uncertainty of the EV diffusion process, namely the gradual adoption of EVs over time, presents notable challenges in emerging markets due to limited data and information. 
Current FCSP studies typically capture the adaptive EV adoption either deterministically or through multiple scenarios \cite{wu2017stochastic}, \cite{8708309}, often based on empirical regression considering exogenous factors such as technological advancements and subsidies. 
Such approaches, however, have failed to incorporate the endogenous impact factor of FCSs on EV adoption. 
Specifically, 
EV diffusion presents an indirect network effect: 
while ongoing EV adoption requires the adaptive expansion of charging facilities to meet the growing demand, strategic FCS deployments can reciprocally encourage more customers to adopt EVs by alleviating range anxiety and improving public perception \cite{mahmutougullari2023robust}. 
Recent evidence shown by empirical regression analysis can support this statement. 
For instance, panel data analysis conducted in the U.S. shows that a 10\% increase in the stock of charging stations can lead to an 8\% increase in EV demand 
\cite{li2017market}. Another study uses autoregression analysis 
to reveal a strong causal promotion effect of public FCSs on EV sales \cite{ma2020deployment}. 
Particularly, the availability of high-speed charging has been highlighted as a crucial factor in promoting EV adoption \cite{illmann2020public}, as it can facilitate long-distance EV travel. 
Consequently, it is necessary to consider the endogenous impact of FCS allocation on future EV diffusion to mitigate the risks of underinvestment and security issues. 
This implies that the uncertainty of the EV adoption should be modeled as \textit{decision-dependent uncertainty (DDU)} in the FCSP problem, whose realization can be reshaped by FCS allocation decisions.

Nonetheless, there are limited FCSP studies that have established explicit formulations for the DDU of EV adoption. 
In \cite{zhang2017incorporating}, the dynamics of EV market share are captured in a multi-period FCSP model as a function of charging opportunities on the path.  
In \cite{anjos2020increasing}, the growth function of EV adoption is approximated with a piecewise linear function w.r.t. the nearby FCSs. 
But both studies fail to account for the stochasticity inherent in EV adoption. 
In \cite{mahmutougullari2023robust}, the impact of the FCS allocation on path-specific EV adoption is formulated using a decision-dependent uncertainty set in a robust FRLM model. 
But oversight of power grid security constraints can lead to unreliable or even infeasible strategies. 
Therefore, there is a lack of FCSP models that have explicitly captured the DDU of EV adoption from the TDN's perspective, which hinders a comprehensive understanding of how this crucial interplay 
can impact the performance of planning strategies and the EV diffusion process.

Another salient issue is that the estimation of the EV diffusion process suffers from potential misspecification due to the scarce information in the emerging EV market \cite{shi2021comprehensive}. 
Currently, the majority of FCSP studies use stochastic optimization (SO) and robust optimization (RO) to model the long-term uncertainty of EV charging. 
For instance, the authors in \cite{wu2017stochastic} and \cite{8708309} generate scenarios with fixed probabilities for EV demands in their stochastic FCSP models. 
But overoptimism toward empirical probabilities often leads to poor statistical significance in reality. 
On the other hand, the RO-based FCSP model utilizes worst-case charging demand within uncertainty sets to attain a robust strategy \cite{mahmutougullari2023robust}, \cite{zhou2020robust}. Nevertheless, overlooking distributional information always results in overconservative solutions.
Distributionally robust optimization (DRO) offers an alternative for modeling ambiguous probability distributions \cite{delage2010distributionally}, which has already been applied to several FCSP studies  \cite{xie2018planning}, \cite{zhou2020planning}. Compared to SO and RO, DRO-based models generate decisions based on the worst distribution in a family of candidate distributions, making a better trade-off between robustness and practicality. However, integrating DDU into the DRO framework significantly complicates the modeling and solution procedures. Recently, related topics have garnered great interest in the field of operations research. 
Specifically, both moment and distance-based DRO problems with DDU have been studied in \cite{luo2020distributionally}, but the development of effective algorithms for solving their non-convex reformulations still poses a challenge. 
In \cite{yu2022multistage}, tractable stochastic counterparts with only decision-independent uncertainty (DIU) are derived for the multi-stage DRO model under DDU,  thereby enabling the application of the SDDiP algorithm. Thus, based on identified gaps and pioneering studies, the following outlines the key contributions of this paper:

\begin{itemize}
    \item[1)] 
    To capture the influence of FCS allocations on EV adoption, a decision-dependent EV diffusion function is formulated. Moreover, to effectively quantify the DDUs of the EV diffusion and address misspecified probability distributions, 
    decision-dependent ambiguity sets (DDASs) are established.
    \item[2)] 
    A two-stage decision-dependent distributionally robust FCSP (D$^3$R-FCSP) model is developed.  
    Based on the worst-case probability distributions of EV adoption rates within DDASs,  the model
    robustly determines the locations and capacities of FCSs with on-site DERs, as well as the expansion of DN assets. 
    Multi-period capacitated ACPC (MCACPC) is incorporated to identify feasible FCS locations and service abilities and capture spatio-temporal EV recharging patterns. 
    \item[3)] To address the nonlinear and non-convex formulation, the two-stage D$^3$R-FCSP model is equivalently recast as a single-level mixed-integer linear programming (MILP) by applying strong duality and the linearization technique. This enables Benders decomposition to facilitate the computation. 
    \item[4)] 
    Comprehensive numerical studies are conducted to exhibit the security insights of our D$^3$R-FCSP model and an accelerated EV diffusion pattern.
    Also, 
    the introduction of a new metric can help decision-makers comprehend the monetary implications of incorporating the DDU in EV diffusion. 
\end{itemize}

The remainder of this paper is organized as follows: We first introduce the decision-dependent EV diffusion model and the MCACPC model in Section II. Then, the detailed formulation of the two-stage D$^3$R-FCSP model is presented in Section III. In Section IV, the tractable MILP reformulation is derived for the D$^3$R-FCSP model, as well as evaluation methods. 
In Section V, numerical tests are conducted to demonstrate the efficiency of the proposed approach. Finally, Section VI concludes our study.

\section{Decision-Dependent EV Diffusion and Charging Uncertainties}

The objective of our FCSP model is to economically allocate RES-powered FCSs in highway TNs to adaptively accommodate the EV charging demands. As the spatio-temporal charging demands are jointly influenced by long-term EV adoption rates and short-term charging behaviors, 
this section will cover the detailed modeling process for these multi-scale random factors. 

\subsection{Strategic Uncertainty: EV Adoption Rate}
\subsubsection{Empirical Decision-Dependent EV Diffusion Function}
The EV diffusion process exhibits indirect network effects \cite{li2017market}. 
The red directed lines shown in Fig. \ref{Uncertainty Modeling} gives an overview of how the mutual reinforcement between FCS investments and EV uptake creates a positive feedback loop. 
To better capture this interplay, 
the logistic model is employed, as it is suitable for depicting the EV diffusion process with a relatively static technical level, and incorporating the impact of external factors such as FCS installation \cite{shi2021comprehensive}. The percentage incremental EV adoption rate in a certain path $od$ is formulated as below: 

\begin{figure}
    \centering
    \includegraphics[width=8.3cm,height=7cm]{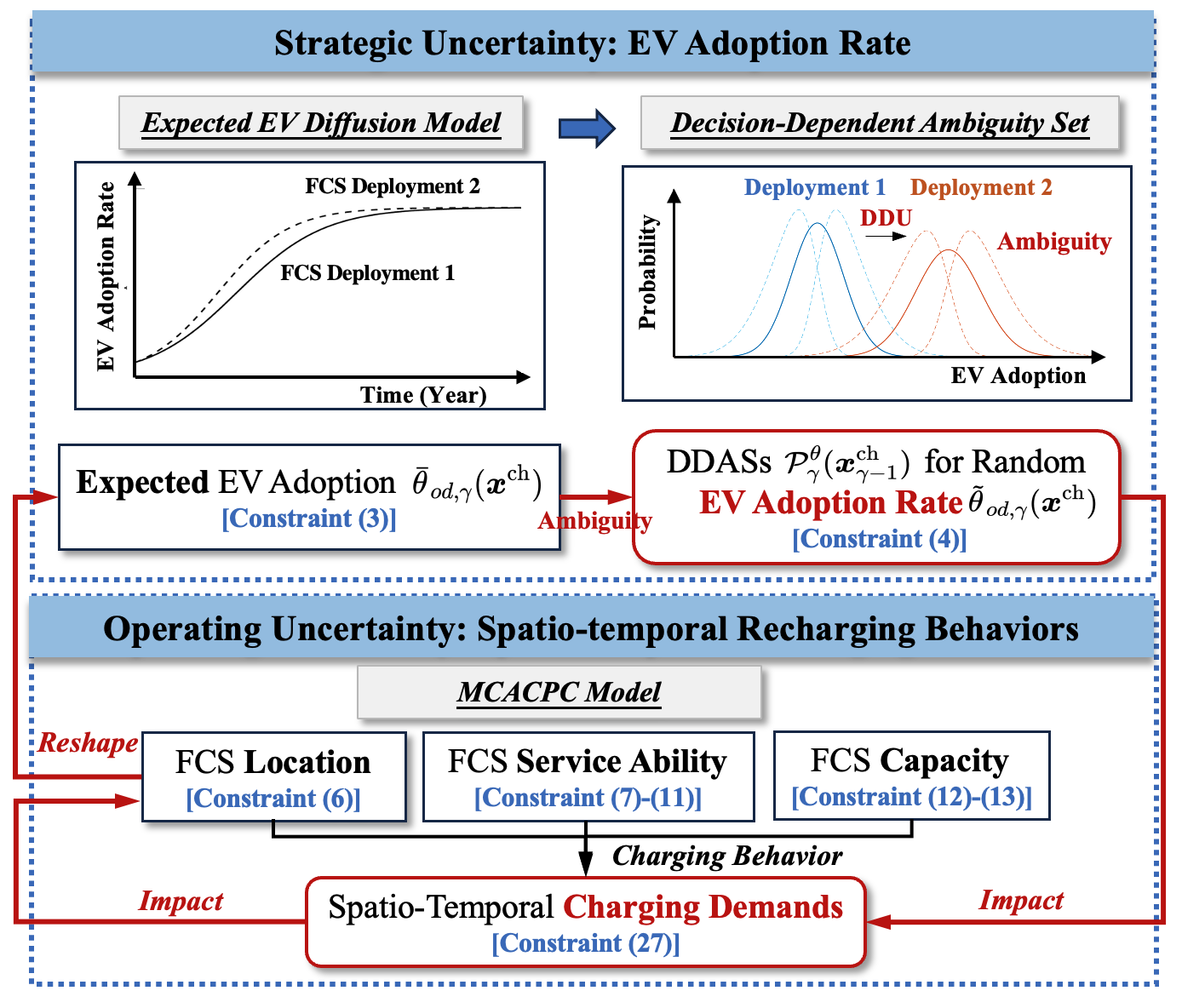}
    \caption{Modeling procedure for strategic/operating uncertainties and an illustration for the indirect network effects}
    \label{Uncertainty Modeling}
    \vspace{-3ex}
\end{figure}

\vspace{-2ex}
\begin{small}
\begin{align}
    \Delta \bar{\theta}_{od,\gamma+1} (\boldsymbol{x}^{\rm ch}) = a_{od,\gamma}
    d_{od,\gamma} (\boldsymbol{x}^{\rm ch})
    [1 \!\!-\!\!\frac{\bar{\theta}_{od,\gamma}(\boldsymbol{x}^{\rm ch})}{K}],
    \forall od\in\mathcal{Q}
    \label{diffusion}
\end{align}
\end{small}
\vspace{-2ex}

\noindent where $\bar{\theta}_{od,\gamma}(\boldsymbol{x}^{\rm ch})$ represents the expected path-specific EV adoption rate, which can be parameterized by the FCS locations $\boldsymbol{x}^{\rm ch}\!\!=\!\!\{{x}_{i, \gamma}^{\rm ch},i\in\mathcal{N}^{\rm TN},\gamma\in\Gamma\}$. 
$K$ is the market potential. $a_{od,\gamma}$ denotes the basic diffusion rate, which is influenced by various impact factors such as the EV price, subsidies, etc. Since those factors are exogenous, their values can be obtained from regression studies \cite{li2017market}, \cite{ma2020deployment}. 
To explicitly capture the influence of $\boldsymbol{x}^{\rm ch}$, 
$d_{od,\gamma}(\boldsymbol{x}^{\rm ch})$ is introduced. This term represents the charging convenience level, which reflects the accessibility of charging facilities along different paths. The higher its value, the more likely it is for EVs to be adopted along the corresponding path. To facilitate a tractable formulation, we present the first assumption below: 
\begin{itemize}
    \item[\textbf{A1}] 
    For a given OD pair, the presence of FCSs along the shortest path has a significantly greater influence on the EV adoption rate, compared to the capacities of FCSs or the presence of FCSs along alternative longer paths. 
\end{itemize}

This is based on the understanding that, during long-distance travel on less congested highways, drivers are generally reluctant to deviate from the shortest path for recharging purposes. 
Furthermore, empirical evidence suggests that the mere presence of FCSs has a far greater impact on EV adoption than their specific capabilities \cite{sierzchula2014influence}. 
Thereby, in line with \cite{mahmutougullari2023robust}, we employ a piecewise linear function to model the convenience level:

\vspace{-2ex}
\begin{small}
\begin{align}
d_{od,\gamma}(\boldsymbol{x}^{\rm ch}) = 
1+\!\!\!\sum_{i\in\mathcal{N}^{\rm TN}_{od}}
    \!\!\Delta d_{od,i}
    x^{\rm ch}_{i,\gamma-1}
    \label{IF}
\end{align}
\end{small}

\vspace{-1.5ex}
\noindent where $\Delta d_{od,i}$ represents the incentive factor (IF), whose magnitude reflects the perception level of the FCS at TN node $i$ in the previous period to drivers along the OD path $od$. 

Furthermore, 
as the logistic diffusion model formulated as \eqref{diffusion}-\eqref{IF} is highly nonlinear and has recursive characteristics, the second assumption is proposed as follows: 
\begin{itemize}
    \item[\textbf{A2}]
    In the early stage of the EV market, we have $\bar{\theta}_{od,\gamma}\ll K$. So we can approximate $\frac{\bar{\theta}_{od,\gamma}(\boldsymbol{x}^{\rm ch})}{K}$ as $\frac{{\theta}_{od,0}
    \cdot \prod_{r=1}^{\gamma-1}a_{od,r}}{K}$, where ${\theta}_{od,0}\neq 0$
represents the initial EV adoption rate at the beginning of the planning horizon.
\end{itemize}

Thus, with the assumption A1 and A2, the expected EV adoption rate can be approximated as follows: 

\vspace{-2ex}
\begin{small}
\begin{align}
    \bar{\theta}_{od,\gamma}(\boldsymbol{x}^{\rm ch}) = 
    \underbrace{
    {\big(\theta_0 + \sum_{r=0}^\gamma 
    a_{od,r} - \sum_{r=1}^\gamma\frac{{\theta}_{od,0}
    \cdot \prod_{r=1}^{\gamma}a_{od,r}}{K}\big)}
    }_{\text{ $\displaystyle \bar{\mu}_{od,\gamma}$}} +
    \notag
    \\
    +\sum_{r=1}^{\gamma}
    \sum_{i\in\mathcal{N}^{\rm TN}_{od}}
    \Big[
    \underbrace{
    \big(a_r-\frac{{\theta}_{od,0}
    \cdot \prod_{r=1}^{\gamma}a_{od,r}}{K}\big)\Delta d_{od,i}
    }_{\Delta \mu_{od,i,\gamma}}
    \cdot x^{\rm ch}_{i,\gamma-1}
    \Big]
    \label{Linearized diffusion}
\end{align}    
\end{small}

\vspace{-1ex}
\noindent which is a piecewise linear function in $\boldsymbol{x}^{\rm ch}$ with the coefficient $\bar{\mu}_{od,\gamma}$ and $\Delta\mu_{od,i,\gamma}$. This  
formulation strikes a balance between accuracy and computational complexity. 
Notably, while the convenience level \eqref{IF} is only parameterized by FCS locations in the last period, 
the EV diffusion process \eqref{Linearized diffusion} is jointly impacted by the EV deployments from the start to the previous period, i.e., ${x}^{\rm ch}_{i,r}\ (r=1,..,\gamma-1)$. This implies that the timing and sequence of FCS allocation potentially play a crucial role in reshaping the EV diffusion trajectory, as visualized in the upper left of Fig. \ref{Uncertainty Modeling}. 

\subsubsection{Decision-Dependent Ambiguity Sets}
Another non-negligible issue is that the lack of information in the emerging EV market can lead to misspecified estimations of future EV adoption. 
Thus, on the basis of the empirical formulation \eqref{Linearized diffusion}, 
period-wise DDASs are constructed to encompass all candidate probability distribution functions (PDFs) for the uncertain EV adoption rates $\tilde{\theta}_{od,\gamma}$:

\vspace{-2.5ex}
\begin{small}
\begin{subequations}
\begin{align}
    &\forall \gamma\!\in\!\Gamma: \mathcal{P}^\theta_{\gamma}(\boldsymbol{x}^{\rm ch})\!=\!
    \big\{ \boldsymbol{\pi}_ \gamma^{\rm st}\!\in\!\mathbb{R}^{|\mathcal{D}_\gamma|}_+,\forall od\in\mathcal{Q}\ |
    \notag
    \\
    &
    \qquad\qquad\qquad
    \sum_{s\in\mathcal{S}_\gamma} \pi^{{\rm st},s}_{\gamma}\! =\! 1,
    \label{DDAS-1}
    \\
    &\bar{\theta}_{od,\gamma}(\boldsymbol{x}^{\rm ch})
    \!-\!\varepsilon_{od,\gamma}^{\mu}
    \!\! \le\!\! \sum_{s\in\mathcal{S}_\gamma}\pi_{\gamma}^{{\rm st},s} \theta_{od,\gamma}^s \!\le\!
    \bar{\theta}_{od,\gamma}(\boldsymbol{x}^{\rm ch})
    \!+\!
    \varepsilon_{od,{\gamma}}^{\mu},
    \label{DDAS-2}
    \\
    &\bar{\upsilon}_{od,\gamma}(\boldsymbol{x}^{\rm ch})\check{\varepsilon}_{od,\gamma}^{\upsilon}
    \!\le\! \sum_{s\in\mathcal{S}_\gamma}\pi_{\gamma}^{{\rm st},s}(\theta_{od,\gamma}^s)^2 \!\le\!
    \bar{\upsilon}_{od,\gamma}(\boldsymbol{x}^{\rm ch})
    \hat{\varepsilon}_{od,{\gamma}}^{\upsilon}\big\}
    \label{DDAS-3}
\end{align}
\label{Decision-dependency}
\end{subequations}
\end{small}

\vspace{-2ex}
\noindent where $\boldsymbol{\theta}^s_\gamma=\{\theta_{od,\gamma}^s,\forall od\in\mathcal{Q}\}$ is the vector for the $s$-th possible realization of EV adoption rates at the $\gamma$-th period. 
$\boldsymbol{\pi}_\gamma=\{\pi_{\gamma}^{{\rm st},s}\}_{s=1}^ {|\mathcal{S}\gamma|}$ represents the probabilities assigned to different realizations, whose total sum should be equal to 1 as the constraint \eqref{DDAS-1} state. 
Specifically, to address the ambiguous estimation 
of path-specific EV adoption rates $\tilde {\theta}_{od,\gamma}$, 
confidence intervals are assigned for its first and second moments, 
as shown in \eqref{DDAS-2} and \eqref{DDAS-3}, respectively. 
Parameters $\varepsilon^\mu_{od,\gamma}$, $\check{\varepsilon}^{\upsilon}_{od,\gamma}$ and $\hat{\varepsilon}^{\upsilon}_{od,\gamma}$ determine the robustness of the DDASs.  
Particularly, if perfect knowledge on the $\mu_{od,\gamma}$ and $\upsilon_{od,\gamma}$ is known, then $\varepsilon^\mu_{od,\gamma}=0$ and $\check{\varepsilon}^{\upsilon}_{od,\gamma}=\hat{\varepsilon}^{\upsilon}_{od,\gamma} = 1$. Otherwise, these values can be adjusted based on historical data, 
risk-aversion levels, etc. 
Here, we have an underlying assumption that every PDF $\mathbb{P}_\gamma$ in the DDAS $\mathcal{P}_{\gamma}^\theta(\boldsymbol{x}^{\rm ch})$ has a decision-independent and period-wise independent finite support $\Xi^\theta_\gamma:= \{\boldsymbol{\theta}_\gamma^s\}_{s=1}^{|\mathcal{S}_\gamma|}$ for all feasible FCS allocation decision $\boldsymbol{x}^{\rm ch}$. 
$\bar{\upsilon}_{od,i}(\boldsymbol{x}^{\rm ch})$ in \eqref{DDAS-3} represents the expected second moment of $\tilde \theta_{od,\gamma}$, and it is utilized to mitigate unrealistic distribution dispersion and prevent overly conservative results:

\vspace{-2.5ex}
\begin{small}
\begin{align}
    \bar{\upsilon}_{od,\gamma}(\boldsymbol{x}^{\rm ch})=
    (\bar{\mu}_{od,\gamma}^2 +\bar{\sigma}_{od,\gamma}^2) (1+\!\!\!\!\sum_{i\in\mathcal{N}^{\rm TN}_{od}}\!\!
    \Delta \upsilon_{od,i,\gamma}
    x^{\rm ch}_{i,\gamma-1})
\label{DDU} 
\end{align}
\end{small}

\vspace{-1ex}
\noindent where $\bar{\sigma}_{od,\gamma}$ is the empirical standard deviation of the EV adoption rate. 
$\Delta \upsilon_{od,i,\gamma}$ is the extent to which the FCS's presence at TN node $i$ affects the variation of EV adoption rate. 
Similar to IF $\Delta d_{od,i}$, their values can be obtained from empirical studies \cite{van2022technology}.

Thus, DDASs are constructed through linear constraints \eqref{Linearized diffusion}-\eqref{DDU}, allowing for the simultaneous consideration of the DDU and the ambiguity of the diffusion estimation. In Section III, DDASs will be incorporated into the D$^3$R-FCSP to identify the worst-case PDFs and ensure a robust strategy. 
Compared to static ambiguity sets adopted in traditional DRO frameworks, the proposed DDASs change with decision variables $\boldsymbol{x}^{\rm ch}$, as intuitively illustrated in the upper right of Fig. \ref{Uncertainty Modeling}. 
The ensuing computational burdens will be discussed and addressed in Section IV. 

\subsection{Operating Uncertainty: EV Charging Demands}
\subsubsection{Assumptions on the EV Charging Logic}
After quantifying the long-term uncertain EV diffusion patterns, it is essential to understand the short-term recharging behaviors in order to make informed decisions regarding FCS location and capacity. 
Without loss of generality, the following assumptions regarding the EV charging logic are made as below:
\begin{itemize}
    \item EVs begin their journey with a fully charged battery, 
    as these nodes typically represent cities and residential areas with type 1 or type 2 charging facilities; 
    \item Journeys are round trips, and the traffic flows follow the shortest paths, which is a reasonable assumption for highways with relatively low traffic density;
    \item Only TN nodes are considered as candidate FCS locations; 
    \item EV drivers have perfect knowledge of the occupancy and vacancies of FCSs.
\end{itemize}

Notably, while these assumptions are made for computational convenience, 
our formulation can be readily adapted to other logic without impacting our solution methodology.

\subsubsection{Multi-Period Capacitated Arc Cover-Path Cover Model}
The ACPC location model is a computationally efficient variant of the FRLM to identify candidate locations for FCSs \cite{abdalrahman2019pev}. 
In our study, the MCACPC model is adopted \cite{zhang2017incorporating}, which additionally considers multi periods and the FCS service ability. 

\begin{figure}
    \centering
    \includegraphics[width=6cm,height=1cm]{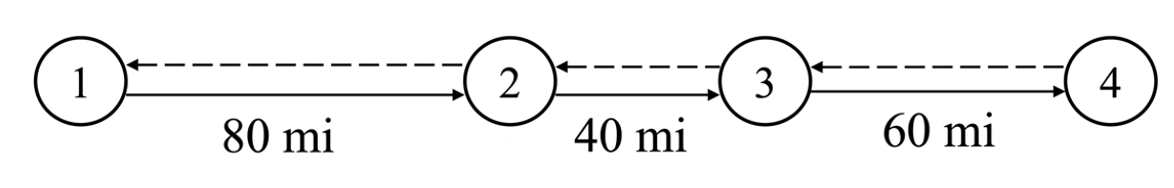}
    \caption{A 4-node transportation network}
    \label{Toy}
    \vspace{-3ex}
\end{figure}

Considering the limited range of EVs, the fundamental logic of the MCACPC 
is that all arcs on all paths should be traversable by the EVs without depleting their battery on the road. 
Specifically, let $\mathcal{A}_{od}^{\rm TN}$ be the set of ordered TN nodes on the shortest path for an OD pair $od$. For each arc $(j,k)$ of $od$, we define the set $\mathcal{K}_{od}(j,k)$ containing all candidate FCS nodes that allow an EV to traverse the arc $(j,k)$ without depleting its battery before reaching node $k$. 
In Fig. \ref{Toy}, assuming the EV driving range is $D^{\rm ev} = 100$ miles, a driver traveling along the OD pair $(1,4)$ who wants to traverse  the arc $(3,4)$ without running out of battery must recharge the battery either at node $2$ or $3$. Therefore, we have $\mathcal{K}_{od}(3,4) = \{2,3\}$. Similarly, to cover the backward path $(3,2)$, the driver must recharge at node $4$ or $3$, so $\mathcal{K}_{od}(3,2) = \{3,4\}$. 
This process is repeated for each arc on each path. 
Algorithm \ref{algorithm 1} in Appendix \ref{ACPC-algo} presents the pseudocode for generating $\mathcal{K}_{od}(j,k)$. 
Then, based on $\mathcal{K}_{od}(j,k)$, the constraints for determining FCS locations can be formulated as follows:

\vspace{-2ex}
\begin{small}
\begin{subequations}
\begin{align}
    &\sum_{i\in\mathcal{K}_{od}(j,k)}\!\!\!
     x^{\rm ch}_{i,\gamma} \ge 1
    \quad 
    \forall od\in \mathcal{Q},
    \gamma\in\Gamma
    \label{location}
    \\
    &x_{i,\gamma}^{\rm ch} \ge 
    x_{i,\gamma-1}^{\rm ch}
    \quad
    \forall i\in \mathcal{N}^{\rm TN},
    \gamma\in\Gamma
    \label{sequence}
    \\
    &\hat{x}_{i,\gamma}^{\rm ch} \ge x_{i,\gamma}^{\rm ch}-x_{i,\gamma-1}^{\rm ch}
    \quad
    \forall i\in \mathcal{N}^{\rm TN},
    \gamma\in\Gamma
    \label{addition}
\end{align}
\label{TN_1}
\end{subequations}
\end{small}
\indent In this formulation, each arc $(j,k)$ on path $od$ provides one constraint of \eqref{location}, ensuring coverage of $(j,k)$ by one of the open facilities. 
Constraints \eqref{sequence} state that installed FCSs will not be abolished in later periods. Constraints \eqref{addition} impose that the installation variable $\hat{x}^{\rm ch}_{i,\gamma}$ is set to 1 only when a new FCS is installed at node $i$ at the begining of period $\gamma$. 

Moreover, it is essential to determine the service ability of an FCS, which represents the maximum number of EVs that can be served concurrently. This is enabled by calculating the spatio-temporal EV recharging patterns as follows: 
\vspace{-2ex}

\begin{small}
    \begin{align}
    \tilde{\Lambda}^{\rm ev}_{od,\gamma,d,t}&(\boldsymbol{x}^{\rm ch}) = 
    \tilde{\theta}_{od,\gamma}(\boldsymbol{x}^{\rm ch})
    {\Lambda}_{od,\gamma,d,t}
    \label{EV flow}
    \\
   \lambda_{i,\gamma,d,t} 
     + & \lambda^{\rm un}_{i,\gamma,d,t}
    =\!\! \sum_{od\in\mathcal{Q}_i}
    \tilde{\Lambda}^{\rm ev}_{od,\gamma,d,t} (\boldsymbol{x}^{\rm ch})fr_{od,i,\gamma,d,t}
    \label{station flow}
    \\ 
    \lambda_{i,\gamma,d,t} & \ge  0, \quad \lambda^{\rm un}_{i,\gamma,d,t}\ge 0
    \label{unsatisfied EV}
    \\
    \forall &
    i\in\mathcal{\mathcal{N}^{\rm TN}}, \gamma\in\Gamma,
    d\in\mathcal{D},
    t\in\mathcal{T}
    \notag
    \\
    &\sum_{i \in \mathcal{K}_{od}(j,k)} \!\!\!
    fr_{od,i,\gamma,d,t} = 1
    \label{total fraction}
    \\
    &0 \le fr_{od,i,\gamma,d,t} \le x^{\rm ch}_{i,\gamma}
    \label{fraction}
    \\
    \forall \gamma\in\Gamma, &
     od\in \mathcal{Q}, (j,k)\in \mathcal{A}^{\rm TN}_{od}, d\in\mathcal{D},
    t\in\mathcal{T}
    \notag
\end{align}
\vspace{-4ex}
\begin{align}
    \lambda_{i,\gamma,d,t}
    \le
    g(z^{\rm ch}_{i,\gamma}) 
    \qquad 
    \forall i\in\mathcal{N}^{\rm TN},
    \gamma\in\Gamma,d\in\mathcal{D},t\in\mathcal{T}
    \label{service ability}
\end{align}
\vspace{-4ex}
\begin{subequations}
    \begin{align}
     &\check{Z}_i^{\rm ch}
     x_{i,\gamma}  \le z^{\rm ch}_{i,\gamma} \le \hat{Z}^{\rm ch}_{i} x_{i,\gamma}
     \qquad
     \forall i\in\mathcal{N}^{\rm TN},
     \gamma\in\Gamma
     \label{CP1}
     \\
     &z^{\rm ch}_{i,\gamma} \ge 
    z^{\rm ch}_{i,\gamma-1}
    \qquad\qquad
     \forall i\in\mathcal{N}^{\rm TN},
     \gamma\in\Gamma
     \label{CP2}
    \\
    & \hat{z}^{\rm ch}_{i,\gamma}
     = z^{\rm ch}_{i,\gamma}
    - z^{\rm ch}_{i,\gamma-1}
    \qquad
     \forall i\in\mathcal{N}^{\rm TN},
     \gamma\in\Gamma
     \label{CP3}
    \end{align}
    \label{CP}
\end{subequations}
\end{small}
\indent In equations \eqref{EV flow}, we define the path-specific EV traffic flows, which are jointly affected by the adoption rate and the daily traffic pattern and thus are also DDUs. 
Notably, to account for the weekly and seasonal variations as well as the daily time-series fluctuations,  traffic flows in representative days with hourly resolution are considered, denoted as $\Lambda_{od,\gamma,d,t}$.  
This strikes a balance between time resolution and tractability. 
Constraints \eqref{station flow} enforce that the total number of EVs entering FCS $i$ during $t$ must be equal to the number of EVs choosing to charge at $i$ across all paths. 
In \eqref{unsatisfied EV}, the unsatisfied EV flows $\lambda^{\rm un}_{i,d,t}$ are permitted and will incur penalties in the objective function. 
Constraints \eqref{total fraction} ensure that all EVs have selected a feasible FCS from the set $\mathcal{K}_{od}(j,k)$ to traverse the arc $(j,k)$. 
Constraints \eqref{fraction} specify that EVs can only get charged at TN nodes with installed FCSs. 
To guarantee service quality and prevent excessive waiting times, the service ability of each FCS is enforced by \eqref{service ability}.  To improve the practicality, we assume that each FCS follows an $M_1/M_2/s$ queue system with a first-come-first-served criterion, and approximate $g(z^{\rm ch}_{i,\gamma})$ as a piece-wise linear function. For more details of the parameter derivation process, readers can refer to \cite{zhang2016pev}. 
Constraints \eqref{CP1} enforce the maximum number of charging spots (CSs) in a single FCS. Constraints \eqref{CP2} state that the CSs installed in the previous period cannot be abolished. \eqref{CP3} defines the integer variable to represent the number of newly installed CSs at the start of period $\gamma$. 
Thus, by capturing the spatio-temporal recharging patterns through the MCACPC model as
\eqref{TN_1}-\eqref{CP}, the eligibility of FCS locations and service abilities can be efficiently enforced. 

\section{Two-Stage Multi-Period D$^3$R-FCSP Model}

\subsection{Planning Framework}
This section presents the formulation for the D$^3$R-FCSP model. The proposed model adopts a two-stage, tri-level structure to adaptively accommodate the decision-dependent EV charging demands, as illustrated in Fig. \ref{Planning framework}. Specifically, the upper level constitutes the first stage of the model for deriving strategic planning decisions, i.e., locations and capacities of FCSs with on-site DERs in the TN and the expansion schedules of the DN. 
In the second stage, a decision-dependent DRO-based bi-level ``max-min" program is nested. The middle level focuses on identifying the worst-case distribution $\mathbb{P}_\gamma^{\rm worst}(\boldsymbol{\tilde{\theta}}_\gamma=\boldsymbol{\theta}_\gamma^s)={\pi_\gamma^{s,{\rm worst}}}$ for EV adoption rates within the DDAS $\mathcal{P}^{\theta}_\gamma(\boldsymbol{x}^{\rm ch})$. Based on $\mathbb{P}_\gamma^{\rm worst}(\boldsymbol{\tilde{\theta}}_\gamma=\boldsymbol{\theta}_\gamma^s)$,  recourse decisions for the daily TDN operation are derived at the lower level. 
Through this framework, the incentive effect of first-stage FCS allocation on the second-stage EV charging demands is explicitly embedded through the middle-level DDASs, which assures a more robust and informed strategy. 
\begin{figure}
    \centering
    \includegraphics[width=8cm,height=9.5cm]{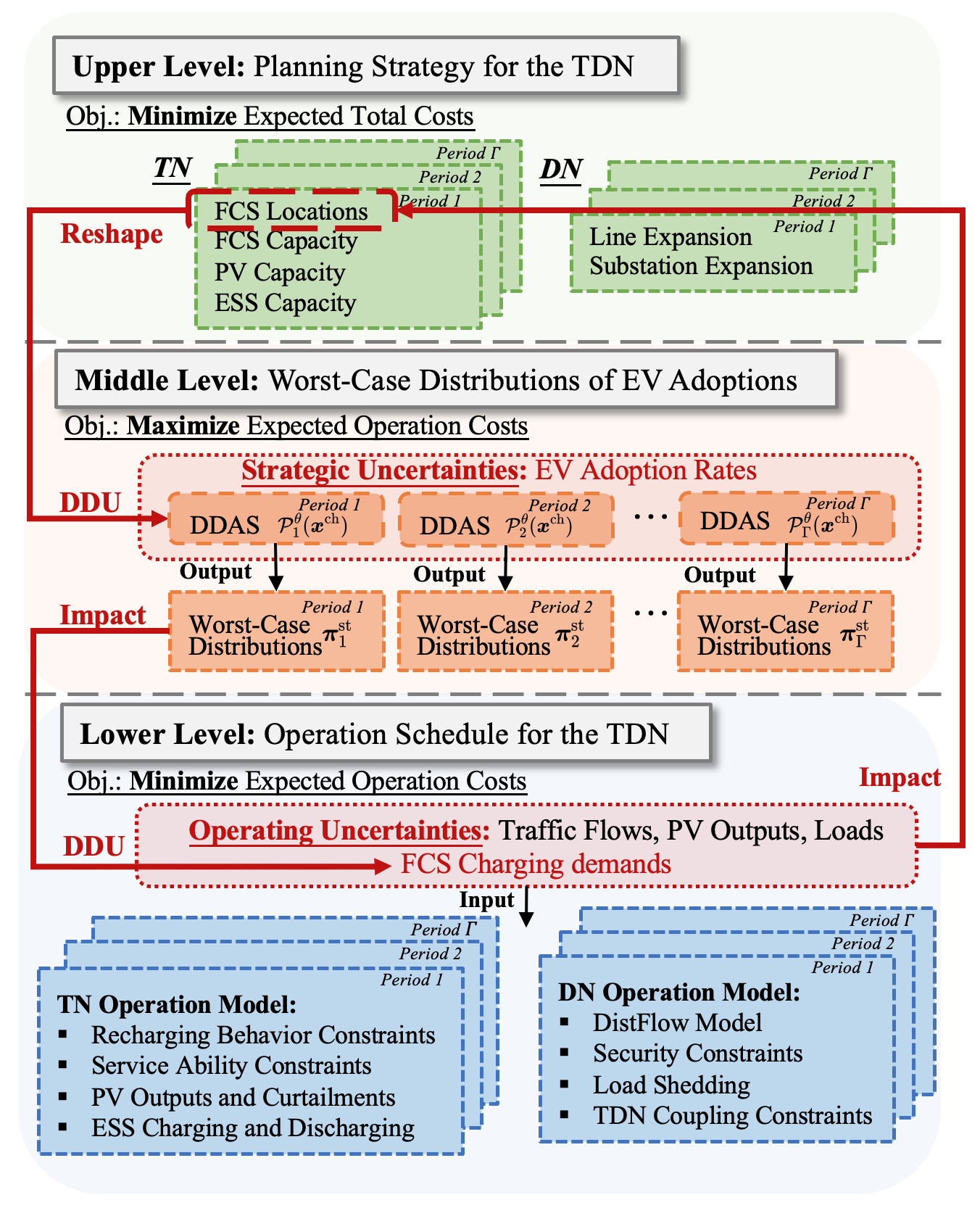}
    \caption{Two-stage tri-level framework for the D$^3$R-FCSP model}
    \label{Planning framework}
\end{figure}
 \vspace{-3ex}
\subsection{Objective Function}
Here, we adopt the perspective of a central planning entity aiming to minimize the overall costs related to investments and expected operation costs: 

\vspace{-1.5ex}
\begin{subequations}
\begin{small}
\begin{align}
    {\mathop{\rm min}_{\boldsymbol{x}_\gamma, \boldsymbol{z}_\gamma}}
    \Big\{
    \sum_{\gamma\in\Gamma}
    \big\{
    c_\gamma^{\rm st}(\boldsymbol{x}_\gamma,\boldsymbol{z}_\gamma)
    + \!\!\!\!\!\!\!\!\!
    {\mathop{\rm max}_{\mathbb{P}_\gamma \in 
    \mathcal{P}_\gamma
    (\boldsymbol{x}_{\gamma-1}^{\rm ch})} }
    \!\!\!\!\!\!
    \mathbb{E}_{\mathbb{P}_\gamma}
    [
    \varphi_\gamma (\boldsymbol{x}_\gamma, \boldsymbol{z}_\gamma, \tilde{\boldsymbol{\theta}}_\gamma)
    ] 
    \big\}
    \Big\}
    \label{first-stage obj}
\end{align}
\end{small}
\noindent \!\!\!\! where $c_\gamma^{\rm st}(\boldsymbol{x}_\gamma,\boldsymbol{z}_\gamma)$ is the first-stage investment cost of the TDN at period $\gamma$. The mid-level maximization aims to find the worst-case distributions $\mathbb{P}^{\rm worst}_\gamma$ in DDASs that lead to the highest cost, thereby attaining robustness. 
The inner minimization problem  $\varphi_\gamma (\boldsymbol{x}_\gamma, \boldsymbol{z}_\gamma, \tilde{\boldsymbol{\theta}}_\gamma)$ is to minimize the overall operation costs:

\vspace{-2.5ex}
\begin{small}
\begin{align}
    \varphi_\gamma (\boldsymbol{x}_\gamma, \boldsymbol{z}_\gamma, \tilde{\boldsymbol{\theta}}_\gamma) \!= \!
    {\mathop{\rm min}_{\boldsymbol{y}_\gamma} }
    \ c_\gamma^{\rm op}(\boldsymbol{y}_\gamma)
    \label{second-stage obj}
\end{align}
\end{small}
\vspace{-3ex}

\noindent where $c_\gamma^{\rm op}(\boldsymbol{y}_\gamma)$ in \eqref{second-stage obj} is the second-stage operation cost of the TDN at the $\gamma$-th planning period. The detailed formulations for $c_\gamma^{\rm st}(\boldsymbol{x}_\gamma,\boldsymbol{z}_\gamma)$ and $c_\gamma^{\rm op}(\boldsymbol{y}_\gamma)$ are presented below:

\vspace{-2.5ex}
\begin{small}
    \begin{align}
    c_\gamma^{\rm st} (\boldsymbol{x}_\gamma,\boldsymbol{z}_\gamma) =&
    (\frac{1}{1+\nu})^{(\gamma-1)}
    \!\left[ic_{\gamma}^{\rm TN}(\boldsymbol{x}_\gamma,\boldsymbol{z}_\gamma)
    \!+\!
    ic_{\gamma}^{\rm DN}
(\boldsymbol{x}_\gamma,\boldsymbol{z}_\gamma)\right]
    \label{C_inv}
    \\
    c_\gamma^{\rm op} (\boldsymbol{y}_\gamma) &=
    \frac{1\!-\!(1\!+\!\nu)^{\gamma}}{\nu}
    \left[oc_{\gamma}^{\rm TN}(\boldsymbol{y}_\gamma)
    +
    oc_{\gamma}^{\rm DN}
    (\boldsymbol{y}_\gamma)\right]
    \label{C_op}
    \\
    {\rm ic}_\gamma^{\rm TN}(\boldsymbol{x}_\gamma,\boldsymbol{z}_\gamma) &= \!\!\!\!
    \sum_{i\in\mathcal{N}^{\rm TN}}(
    a_\gamma^{\rm st}C^{\rm in}\hat{x}^{\rm ch}_{i,\gamma} +
    a_\gamma^{\rm ch} C^{\rm ch}\hat{z}^{\rm ch}_{i,\gamma} + \notag
    \\
    & a_\gamma^{\rm re}C^{\rm re}\hat{P}^{\rm re}_{i,\gamma} + a_\gamma^{\rm es} C^{\rm es} \hat{E}^{\rm es}_{i,\gamma} ) 
    \label{C_inv_tn}  
    \\
    {\rm ic}_\gamma^{\rm DN}(\boldsymbol{x}_\gamma,\boldsymbol{z}_\gamma) &= \!\!\!\!\!\!\!
    \sum_{(n,m)\in\mathcal{\rm L}}\!\!\!\!\!
    a^{\rm L}_\gamma C^{\rm L}L_{nm}\bar{P}^{\rm L}\hat{x}_{nm,\gamma}^{\rm L}\!
    +\!\!\!\!\!
    \sum_{i\in\mathcal{N}^{\rm TN}}\!\!\!a_\gamma^{\rm sub}C^{\rm sub}\hat{P}^{\rm sub}_{i,\gamma}
    \label{C_inv_dn}
    \\
    {\rm oc_\gamma^{\rm TN}}(\boldsymbol{y}_\gamma) &= \!\!\!\!
    \sum_{i\in\mathcal{N}^{\rm TN}}
    \!
    \sum_{d\in\mathcal{D}_\gamma}
    \!
    \sum_{t\in\mathcal{T}}
    W_{d}  C^{\rm un}
    \lambda_{i,d,t}^{\rm un} 
    \label{C_op_tn}
    \\
    {\rm oc_\gamma^{\rm DN}}(\boldsymbol{y}_\gamma) &=
    \!\!\!\!
    \sum_{n\in\mathcal{N}^{\rm DN}}
    \!
    \sum_{d\in\mathcal{D}_\gamma}
    \!
    \sum_{t\in\mathcal{T}_{d}}
    W_{d} ( C^{\rm up,p}p^{\rm up}_{n,d,t}
    \notag
    \\
    &\qquad\qquad
    +C^{\rm cu}p^{\rm re,cu}_{n,d,t}+
    C^{\rm sh}
    \zeta_{n,d,t} P^{\rm load}_{n,d,t})
    \label{C_op_dn}
    \end{align}
\end{small}
\label{obj}
\end{subequations}
\indent The investment cost of TN 
$ic_\gamma^{\rm TN}$ in \eqref{C_inv_tn} encompasses four terms: the installation cost of new FCSs, CSs, on-site PVs, and on-site ESSs. The investment cost of DN 
$ic_\gamma^{\rm DN}$ in \eqref{C_inv_dn} comprises the replacement cost of line conductors and the expansion cost of substations. 
The operation cost of the TN 
$oc_\gamma^{\rm TN}$ in \eqref{C_op_tn} includes the penalty cost for unsatisfied EV charging demands. 
The operation cost of DN
$oc_\gamma^{\rm DN}$ in \eqref{C_op_dn} consists of the energy purchase cost from the upstream main grid, the penalty cost of PV curtailment and load shedding. 
Both investment and operation costs are converted to the present value using the interest rate $\nu$, as the coefficients in \eqref{C_inv} and \eqref{C_op} show. 
Additionally, $a_\gamma^{\rm \cdot}$ in \eqref{C_inv_tn} and \eqref{C_inv_dn} represent the capital recovery factors for different resources to FCSs, CSs, ESSs, PVs, line conductors and substations, which convert the present investment costs at period $\gamma$ into equal annual payments over their respective lifespans.  


\subsection{Upper Level: \!RES-Powered\! FCS Allocation\! and\! DN\! Expansion}
In addition to the location and capacity decisions for FCSs constrained by \eqref{TN_1} and \eqref{CP} in Section II, the upper-level optimization also determines the size of on-site PVs and ESSs in FCSs, as well as the expansion of lines and substations in the DN to avoid congestion during the operation:  

\vspace{-3ex}
\begin{small}
\begin{subequations}
    \begin{align}
    P^{\rm re}_{i,\gamma} & = \!\!\! \sum\nolimits_{r=1}^\gamma 
    \!\!\!
    \hat{P}^{re}_{\i,r}, \  
    E^{\rm es}_{i,\gamma} = 
    \!\!\!\sum\nolimits_{r=1}^\gamma \!\!\! \hat{E}^{\rm es}_{i,r},\quad 
    \forall i\in\mathcal{N}^{\rm TN}, \gamma\in\Gamma
    \label{Installed PVESS}
    \\
    0  \le & {P}^{\rm re}_{i,\gamma} \le \overline{P}^{\rm re}_{i} x_{i,\gamma},
    0 \le {E}^{\rm es}_{i,\gamma} \le \overline{E}^{\rm es}_{i} x_{i,\gamma},
    \forall i\in\mathcal{N}^{\rm TN},\gamma\in\Gamma
    \quad
    \label{PVESS Limits}
    \\
    &\qquad \hat{P}^{\rm re}_{i,\gamma}\ge 0,\quad
    \hat{E}^{\rm re}_{i,\gamma}\ge 0
    \qquad 
    \forall i\in\mathcal{N}^{\rm TN},\gamma\in\Gamma
    \end{align}
    \label{ESS}
\end{subequations}
\vspace{-5ex}
\begin{subequations}
    \begin{align}
        &x^L_{ij,\gamma}\ge x^L_{ij,\gamma-1}
        \quad
    \forall (i,j)\in\mathcal{L},
    \gamma\in\Gamma
        \\
        &\hat{x}^L_{ij,\gamma}\ge
        x^L_{ij,\gamma} - x^L_{ij,\gamma-1}\quad
        \quad
    \forall (i,j)\in\mathcal{L},
    \gamma\in\Gamma
    \end{align}
    \label{line investment}
\end{subequations}
\vspace{-4ex}
\begin{subequations}
    \begin{align}
        P^{\rm sub}_{i,\gamma} = 
        & P_{i}^{\rm sub,0} + \sum\nolimits_{r=1}^\gamma \hat{P}_{i,r}^{\rm sub}
        \quad
    \forall i\in\mathcal{N}^{\rm TN},
    \gamma\in\Gamma
        \\
        &\hat{P}_{i,\gamma}^{\rm sub}\ge 0
        \quad
    \forall i\in\mathcal{N}^{\rm TN},
    \gamma\in\Gamma
        \\
        &P^{\rm sub}_{i,\gamma} \ge \bar{P}^{\rm ch}{z}^{\rm ch}_{i,\gamma}
        \quad
    \forall i\in\mathcal{N}^{\rm TN},
    \gamma\in\Gamma
    \end{align}
    \label{substation investment}
\end{subequations}
\end{small}
\vspace{-3ex}

\noindent where constraints \eqref{ESS} 
state the installed capacity of PVs and ESSs at each period as well as enforce their maximum capacity. 
Constraints \eqref{line investment} ensure the preservation of new conductors replaced in previous periods and define variables representing the the newly expanded lines. 
Constraints \eqref{substation investment} delineate the required capacity expansion of substations. 
\vspace{-2ex}

\subsection{Middle Level: Worst-Case Distributions for EV Adoption}
At the middle level, the worst-case distributions for EV adoption rates that can result in the maximal operation cost are identified in DDASs. The pertinent constraints are presented by \eqref{Linearized diffusion}-\eqref{DDU} in Section II. 

\vspace{-2ex}
\subsection{Lower Level: Operational Schedule of the TDN}
At the lower level, the traffic flow and EV recharging pattern are formulated by \eqref{EV flow}-\eqref{service ability}. Other constraints are enforced below:


\vspace{-2ex}
\begin{small}
\begin{subequations}
\begin{align}
    &p^{\rm re}_{i,d,t}+p^{\rm re,cu}_{i,d,t} = {\varpi}_{i,d,t}{P}_{i,\gamma}^{\rm re}
    \\ 
    &p^{\rm re}_{i,d,t} \ge 0,\ 
    p^{\rm re,cu}_{i,d,t} \ge 0
\end{align}
\label{WT}
\end{subequations}
\vspace{-5ex}
\begin{subequations}
    \begin{align}
        &0 \le\ p^{\rm es,c}_{i,d,t} \le E^{\rm es}_{i,\gamma}\iota^{\rm c}
        \\
        &0 \le p^{\rm es,d}_{i,d,t} \le E^{\rm es}_{i,\gamma}\iota^{\rm d}
    \end{align}
    \label{ESS charging rate}
\end{subequations}
\vspace{-5ex}
\begin{subequations}
    \begin{align}
        e^{\rm es}_{i,d,t} =& e^{\rm es}_{i,d,t-1} + \eta^{\rm c} p^{\rm es,c}_{i,d,t}\Delta t - (1/{\eta^{\rm d}})p^{\rm es,d}_{i,d,t}\Delta t
        \label{residualenergy}
        \\
        &
        0 \le e^{\rm es}_{i,d,t} \le {E}^{\rm es}_{i,\gamma}
        \label{esslimits}
        \\
        &\forall i\in\mathcal{N}^{\rm TN},\gamma\in\Gamma, d\in\mathcal{D}_\gamma,d,t\in\mathcal{T}
    \notag
    \end{align}
    \label{TN_end}
\end{subequations}
\vspace{-4ex}
\begin{subequations}
\begin{align}
    &(1-\zeta_{n,d,t}) {P}^{\rm load}_{n,d,t} + p_{n,d,t}^{\rm ch} +
    \!\!\!\!\!\!\sum_{m|(m,n)\in\mathcal{L}}\!\!\!\!\!\!p^{\rm line}_{mn,t} -
    \!\!\!\!\!\!\!\!\sum_{m|(n,m)\in\mathcal{L}}\!\!\!\!\!\!p^{\rm line}_{mn,t}
    = 
    \notag\\
    &\qquad\qquad p^{\rm up}_{n,d,t} +
    p^{\rm es,d}_{n,d,t} -
    p^{\rm es,c}_{n,d,t} +
    p^{\rm re}_{n,d,t} 
    \\
    &(1\!-\!\zeta_{n,d,t}) {Q}^{\rm load}_{n,d,t} \!+
    \!\!\!\!\!\!\!\sum_{m|(m,n)\in\mathcal{L}}\!\!\!\!\!\!\!q^{\rm line}_{mn,t} -
    \!\!\!\!\!\!\!\!\sum_{m|(n,m)\in\mathcal{L}}\!\!\!\!\!\!\!q^{\rm line}_{mn,t}
    = 
    q^{\rm up}_{n,d,t} 
    \\
    &\qquad\qquad
    \forall n\in \mathcal{N}^{\rm DN},\gamma\in\Gamma,d\in\mathcal{D}_\gamma,d,t\in\mathcal{T} \notag
\end{align}
\label{powerbalance}
\end{subequations}
\vspace{-4ex}
\begin{align}
    p^{\rm up}_{n,d,t} &\ge 0, q^{\rm up}_{n,d,t} \ge 0
\\
    u^{\rm sqr}_{n,d,t}-u^{\rm sqr}_{m,t} &=2(R_{nm} p^{\rm line}_{nm,t}+X_{nm} q^{ \rm line}_{nm,t})
    \label{DistFlow}
\end{align}
\vspace{-4ex}
\begin{subequations}
\begin{align}
    & -(\overline P_{nm}^{\rm line}+\overline{P}^{\rm L}x^{\rm L}_{nm,\gamma})
    \le p^{\rm line}_{nm,t} \le \overline P_{nm}^{\rm line} +\overline{P}^{\rm L}x^{\rm L}_{nm,\gamma}
     \label{line capacity P}
     \\
    &-(\overline Q_{nm}^{\rm line} +\overline{Q}^{\rm L}x^{\rm L}_{nm,\gamma}) \le q^{\rm line}_{nm,t} \le  \overline Q_{nm}^{\rm line} +\overline{Q}^{\rm L}x^{\rm L}_{nm,\gamma}
    \label{line capacity Q}
    \\
    &\qquad \qquad \forall (n,m) \in \mathcal{L},\gamma\in\Gamma,d\in\mathcal{D}_\gamma, t\in\mathcal{T}
    \notag
\end{align}\label{line capacity}
\end{subequations}
\vspace{-4ex}
\begin{align}
    \underline{U}_n^{\rm sqr} \le u^{\rm sqr}_{n,d,t}\le  \overline U_n^{\rm sqr}&,
    \forall n \in \mathcal{N}^{\rm DN},
    \gamma\in\Gamma,d\in\mathcal{D}_\gamma, t\in\mathcal{T}
    \label{voltage}
    \\
    0\le \zeta_{n,d,t}
    \le  1 &,
    \forall n \in \mathcal{N}^{\rm DN},
   \gamma\in\Gamma, d\in\mathcal{D}_\gamma, t\in\mathcal{T}
    \label{load shedding}
    \\
     p_{n,d,t}^{\rm ch}\!=
    \!\!\!\!
    \sum_{i\in\mathcal{N}_n^{\rm T\text{-}D}}
    \!\! 
    \frac{Ed\cdot D^{\rm ev}}{\eta^{\rm ev}}&
    \lambda_{i,d,t},
    \forall i\!\in\!\mathcal{N}_n^{\rm T\text{-}D}\!\!,\gamma\in\Gamma, d\!\in\!\mathcal{D}_\gamma,t\!\in\!\mathcal{T}
    \label{coupling}
\end{align}
\end{small}  
\vspace{-2ex}

To attain a tractable formulation, the second-stage random variables $\{\tilde{\Lambda}_{od,\gamma},\tilde{\varpi}_{i,\gamma}, \tilde{P}^{\rm load}_{n,\gamma}\}$ are discretized in to $\{{\Lambda}_{od,d,t},{\varpi}_{i,d,t},{P}^{\rm load}_{n,d,t},d\in\mathcal{D_\gamma},t\in\mathcal{T}\}$ based on different representative days $d\in\mathcal{D}_\gamma$ with different time period $t\in\mathcal{T}_d$. 
Specifically, constraints \eqref{WT} determine the PV outputs, and curtailments are allowed and penalized in the objective function \eqref{C_op_tn}.
Constraints \eqref{ESS charging rate} restrict the charging and discharging rates of the ESSs. The state of charge (SoC) of ESSs is presented by \eqref{TN_end}. 
Constraints \eqref{powerbalance}-\eqref{DistFlow} present the linearized DistFlow formulation \cite{Baran1989Network}. Constraints \eqref{powerbalance} state the nodal power balance. 
Constraints \eqref{DistFlow} represent the voltage drops across distribution lines. 
Line flows and node voltages are restricted within their technical limits, as enforced by \eqref{line capacity} and \eqref{voltage}. Constraints \eqref{load shedding} limit the maximum shed loads. Constraints \eqref{coupling} calculate the average charging demand during $t$ at coupling points of the TDN.

Consequently, the two-stage multi-period D$^3$R-FCSP model is formulated with \eqref{Linearized diffusion}-\eqref{coupling}. 
As highlighted by the directed red lines in Fig. \ref{Planning framework}, not only do the first-stage FCS allocation decisions $\boldsymbol{x}^{\rm ch}$ have a forward effect on second-stage EV charging demands, but $\boldsymbol{x}^{\rm ch}$ are also implicitly affected backward by the reshaped EV charging patterns. The incorporation of this mutual reinforcement will also be examined in the numerical studies in Section V.  
In addition, as the non-convex, non-linear formulation makes direct computation prohibitive, the reformulation and the solution methodology are presented in the subsequent section. 

\section{Solution and Evaluation Methodology}
\subsection{Compact Form}
In this section, we introduce the reformulation, solution and evaluation methods for the D$^3$R-FCSP model. To simplify the exposition, we first present the compact form below:

\vspace{-2ex}
\begin{subequations}
\begin{small}
\begin{align}
    {\mathop{\rm min}_{\boldsymbol{u},\boldsymbol{w}}}
    F^{\rm DDU}(\boldsymbol{u},\boldsymbol{w})=  {\mathop{\rm min}_{\boldsymbol{u}_{\gamma},\boldsymbol{w}_\gamma}}&\Big\{\sum_{\gamma\in\Gamma} 
    \big\{
    (\boldsymbol{c}_{1,\gamma} \boldsymbol{u}_\gamma + \boldsymbol{c}_{2,\gamma} \boldsymbol{w}_\gamma)
    \notag
    \\
     + 
    {\mathop{\rm sup}_{\mathbb{P}_\gamma \in 
    \mathcal{P}_\gamma
    (\boldsymbol{x}_{\gamma-1}^{\rm ch})} }
    \!\!\!\!
    \mathbb{E}_{\mathbb{P}_\gamma}
    \big[
    \varphi_\gamma &(\boldsymbol{u}_\gamma, \boldsymbol{w}_\gamma, \tilde{\boldsymbol{\theta}}_\gamma)
    \big] 
    \big\}
    \Big\}
    \\
    {\text{s.t.}}\quad
    \boldsymbol{A}_\gamma\boldsymbol{u}_\gamma
    &+ 
    \boldsymbol{B}_\gamma\boldsymbol{w}_\gamma
    \le \boldsymbol{b}_\gamma
    \quad
    \forall \gamma\in\Gamma
    \label{compact-couple}
\end{align}
\end{small}
where $\boldsymbol{c}_{1,\gamma}$, $\boldsymbol{c}_{2,\gamma}$ are cost coefficient vectors related to the binary allocation and sizing decisions at period $\gamma$, i.e., $\boldsymbol{u}_\gamma$ and $\boldsymbol{w}_\gamma$, respectively. $\boldsymbol{A}_\gamma$, $\boldsymbol{B}_\gamma$ and $\boldsymbol{b}_\gamma$ in \eqref{compact-couple} are the coefficient matrices and the right-hand side parameter vector for the first-stage constraints, i.e., \eqref{TN_1}-\eqref{CP} and \eqref{ESS}-\eqref{substation investment}. 
The second-stage problem at period $\gamma$ is rewritten as follows:

\vspace{-2ex}
\begin{small}
\begin{align}
    \varphi_\gamma (\boldsymbol{u}_\gamma,\boldsymbol{w}_\gamma,\tilde{\boldsymbol{\theta}}_\gamma ) &= 
   {\mathop{\rm min}_{\boldsymbol{y}_d} }
   \sum_{d\in\mathcal{D}_\gamma}
   W_{d}
   \boldsymbol{c}_{3,\gamma} \boldsymbol{y}_d
   \label{compact-second-objective}
    \\
    {\text{s.t.}}\quad
    \boldsymbol{C}_d \boldsymbol{y}_d \le \boldsymbol{l}_d 
    - &\boldsymbol{D}_d\boldsymbol{u}_\gamma - \boldsymbol{E}_d\boldsymbol{w}_\gamma - \boldsymbol{F}_d\boldsymbol{\tilde{\theta}}_\gamma,
    \forall d\in\mathcal{D}_\gamma
    \label{compact-second-constraint}
\end{align}
\label{compact-instage}
\end{small}
\end{subequations}
\vspace{-2ex}

\noindent where $\boldsymbol{c}_{3,\gamma}$ is the cost coefficient vector for the second-stage operational decision vector $\boldsymbol{y}_d$ in 
\eqref{C_op}. $\boldsymbol{l}_d$ is the right-hand side parameter vector, while $\boldsymbol{C}_d$, $\boldsymbol{D}_d$, $\boldsymbol{E}_d$, and $\boldsymbol{F}_d$ are coefficient matrices in the second-stage constraints \eqref{station flow}-\eqref{fraction} and \eqref{WT}-\eqref{load shedding}. Thus, formulation \eqref{compact-instage} combined with DDASs \eqref{Linearized diffusion}-\eqref{DDU} constitute the compact form of the D$^3$R-FCSP model.

\vspace{-2ex}
\subsection{Reformulation of the D$^3$R-FCSP Model}
Since DDASs for EV adoption rates can be affected by the decisions for FCS locations, $\boldsymbol{x}^{\rm ch}$,  
we first decouple the decision-dependency in the DDAS to achieve tractability.
As unsatisfied EVs, load shedding, and PV curtailment are allowed, 
the formulated model has complete recourse, ensuring that $\varphi(\boldsymbol{x},\boldsymbol{z},\boldsymbol{\theta})$ has an upper bound for any given first-stage decisions $(\boldsymbol{x}_\gamma,\boldsymbol{z}_\gamma)$ and any realization $\boldsymbol{\theta}_\gamma\in \Xi^\theta_\gamma$. This fact allows us to 
derive the following proposition by utilizing the strong duality of the middle level (the detailed proof is presented in Appendix A): 
\begin{proposition}
    \normalfont
If for any feasible first-stage decisions $[\boldsymbol{u}_\gamma,\boldsymbol{w}_\gamma,\forall \gamma\in\Gamma]$, the DDASs defined in \eqref{Linearized diffusion}-\eqref{DDU} is non-empty, then the D$^3$R-FCLS model \eqref{compact-instage} can be reformulated as the below single-level problem \cite{yu2022multistage}:

\vspace{-3ex}
\begin{small}
\begin{subequations}
\begin{align}
    &\min_{
    \boldsymbol{u}_\gamma,
    \boldsymbol{w}_\gamma,
    \boldsymbol{y}_\gamma,
    \boldsymbol{\nu}_\gamma,
    \atop
    \boldsymbol{\kappa}_\gamma,
    \boldsymbol{\alpha}_\gamma\ge 0, \boldsymbol{\beta}_\gamma\ge 0
    } 
    \sum_{\gamma\in\Gamma}\Big\{
    \boldsymbol{c}_{1,\gamma}\boldsymbol{u}_\gamma 
    \!+\!
    \boldsymbol{c}_{2,\gamma}\boldsymbol{w}_\gamma
    \!+\! \kappa_\gamma
    \!+\!\!\!\!
    \sum_{od\in\mathcal{Q}}\!\!
    \big[
    (\varepsilon_{od,\gamma}^{\mu}\!-\!\bar{\mu}_{od,\gamma})
    \alpha^\mu_{ od,\gamma} 
    \notag
    \\
    &-
    \sum_{r=1}^{\gamma-1}
    \sum_{i\in\mathcal{N}_{od}^{\rm TN}} \!\!\Delta \mu_{od,i,\gamma} \nu^{\rm I}_{od,i,\gamma,r}
    \!-\! (\bar{\mu}_{od,\gamma}^{2}\!+\!\bar{\sigma}_{od,\gamma}^{2}) \check{\varepsilon}_{od,\gamma}^{\upsilon}
    \alpha^\upsilon_{od,\gamma}
    \notag
    \\
    & -\!\! \sum_{i\in\mathcal{N}_{od}^{\rm TN}}\!\!\!\! (\bar{\mu}_{od,\gamma}^{2}\!+\!\bar{\sigma}_{od,\gamma}^{2})
    \Delta\upsilon_{od,i,\gamma} \check{\varepsilon}_{od,\gamma}^{\upsilon} \nu^{\rm II}_{od,i,\gamma,\gamma-1} 
    \!+\! (\bar{\mu}_{od,\gamma}\!+\!\varepsilon_{od,\gamma}^{\mu}) \beta^{\mu}_{od,\gamma}
    \notag
    \\
    &\!+ 
    \sum_{r=1}^{\gamma-1}
    \sum_{i\in\mathcal{N}_{od}^{\rm TN}}\!\!\Delta \mu_{od,i,\gamma} \nu^{\rm III}_{od,i,\gamma,r} \!+\! (\bar{\mu}_{od,\gamma}^{2}\!+\!\bar{\sigma}_{od,\gamma}^{2}) \hat{\varepsilon}_{od,\gamma}^{\upsilon}\beta^{\upsilon}_{od,\gamma}
    \notag
    \\
    &\! +\! \sum_{i\in\mathcal{N}_{od}^{\rm TN}} \!\!\!(\bar{\mu}_{od,\gamma}^{2}\! +\!\bar{\sigma}_{od,\gamma}^{2})
    \Delta\upsilon_{od,i,\gamma} \hat{\varepsilon}^\upsilon_{od,\gamma} \nu^{\rm IV}_{od,i,\gamma,\gamma-1}
    \big]
    \Big\}
    \label{reform-obj}
    \\
    &{\text{s.t.:}}\qquad\qquad\qquad
    \boldsymbol{A}_\gamma\boldsymbol{u}_\gamma
    + 
    \boldsymbol{B}_\gamma\boldsymbol{w}_\gamma 
    \le \boldsymbol{b}_\gamma
    \quad
    \forall \gamma\in\Gamma
    \label{ref-con1}
    \\
    &\qquad\qquad\qquad(\nu^{\rm I}_{od,i,\gamma,r},\alpha^\mu_{od,\gamma},x^{\rm ch}_{i,r}) \!\in\! \boldsymbol{M}^I_{od,i,\gamma,r}
    \notag
    \\
    &\qquad\qquad\qquad(\nu^{\rm II}_{od,i,\gamma,r},\alpha^\upsilon_{od,\gamma},x^{\rm ch}_{i,r}) \!\in\! \boldsymbol{M}^{\rm II}_{od,i,\gamma,r}
    \notag
    \\
    &\qquad\qquad\qquad(\nu^{\rm III}_{od,i,\gamma,r},\beta^\mu_{od,\gamma},x^{\rm ch}_{i,r}) \!\in\! \boldsymbol{M}^{\rm III}_{od,i,\gamma,r}
    \notag
    \\
    &(\nu^{\rm IV}_{od,i,\gamma,r},\beta^\upsilon_{od,\gamma},x^{\rm ch}_{i,r}) \!\in\! \boldsymbol{M}^{\rm IV}_{od,i,\gamma,r},
    \forall od\!\in\!\mathcal{Q},i\!\in\!\mathcal{N}^{\rm TN}\!\!,\gamma\!\in\!\Gamma
    \label{ref-con2}
    \\
    & 
    \kappa_{\gamma} + \sum_{{od}\in\mathcal{Q}} \!\theta_{od,\gamma}^s (\beta^{\mu}_{od,\gamma}\!-\!\alpha^{\mu}_{od,\gamma})\! + \!\sum_{od\in\mathcal{Q}}\!(\theta_{od,\gamma}^s)^2(\beta^{\upsilon}_{od,\gamma}\!-\!\alpha^{\upsilon}_{od,\gamma})
    \notag
    \\
    &\qquad \qquad\qquad\qquad
    \ge 
    \sum_{d\in\mathcal{D}_\gamma} W_d\boldsymbol{c}_3\boldsymbol{y}_d^s
    \qquad 
    \forall \gamma \in \Gamma, s\in\mathcal{S}_\gamma 
    \label{ref-dual constraint}
    \\
     &\qquad\qquad \boldsymbol{C}_d \boldsymbol{y}_d^s \le \boldsymbol{l}_d 
    - \boldsymbol{D}_d\boldsymbol{u}_\gamma - \boldsymbol{E}_d\boldsymbol{w}_\gamma - \boldsymbol{F}_d\boldsymbol{{\theta}}^s_\gamma
    \notag
    \\
    &
    \qquad\qquad\qquad\qquad\qquad\qquad
    \forall \gamma\in\Gamma, s\in\mathcal{S}_\gamma, d\in\mathcal{D}_\gamma
\end{align}
\label{Reformulated Problem}
\end{subequations}
\end{small}

\vspace{-2ex}
\!\!\!\!\!\! \noindent where $\kappa_\gamma$, $\boldsymbol{\alpha}_\gamma\!\!=\!\![\alpha^\mu_{od,\gamma},\alpha^\upsilon_{od,\gamma},\forall od\!\in\!\!\mathcal{Q}]^{\rm T}$, and $\boldsymbol{\beta}_\gamma\!\!=\!\![\beta^\mu_{od,\gamma},\beta^\upsilon_{od,\gamma},\!\forall od\!\in\!\mathcal{Q}]^{\rm T}$ are dual variables corresponding to \eqref{DDAS-1}, the left-hand side constraints,  
and the right-hand side constraints of \eqref{DDAS-2} and \eqref{DDAS-3}, respectively. 
$\boldsymbol{y}^s_d$ is the vector of second-stage variables under the $s$-th scenario of EV adoption rate. 
Furthermore, considering the binary nature of the allocation decision $x^{\rm ch}_{i, \gamma}$, we introduce auxiliary variables $\nu^{\rm I}_{od,i,\gamma,r}$, $\nu^{\rm II}_{od,i,\gamma,r}$, $\nu^{\rm III}_{od,i,\gamma,r}$ and $\nu^{\rm IV}_{od,i,\gamma,r}$ to exactly reformulate the bilinear terms $\alpha^{\mu}_{od,\gamma}x^{\rm ch}_{i,r}$, $\alpha^{\upsilon}_{od,\gamma}x^{\rm ch}_{i,r}$, $\beta^{\mu}_{od,\gamma}x^{\rm ch}_{i,r}$, and $\beta^{\upsilon}_{od,\gamma}x^{\rm ch}_{i,r}$ using McCormick envelopes $\boldsymbol{M}^{\rm I}_{od,i,\gamma,r}$, $\boldsymbol{M}^{\rm II}_{od,i,\gamma,r}$, $\boldsymbol{M}^{\rm III}_{od,i,\gamma,r}$, and $\boldsymbol{M}^{\rm IV}_{od,i,\gamma,r}$, as \eqref{ref-con2} show. Details on McCormick envelope method can be found in \cite{mccormick1976computability}. 
\end{proposition}
Proposition 1 enables the exact reformulation of the original D$^3$R-FCSP model into a single-level MILP. 
However, given that our model extensively considers multi-scale uncertainties, the total number of scenario combinations is $|\mathcal{S}\gamma| \times |\Gamma| \times |\mathcal{D}\gamma|$. Furthermore, the linear reformulation based on the McCormick envelope adds extra sets of constraints. 
To address these computational burdens, we implement the Benders decomposition algorithm \cite{geoffrion1972generalized}. Detailed steps for execution can be found in Appendix \ref{app-BD}.

\vspace{-3ex}
\subsection{Worst-Case Distributions and the Evaluation Metric}
\subsubsection{Worst-Case Distributions} 
The explicit derivation of extremal distributions that attain the worst-case expectation within DDASs is essential for assessing the hidden risks of different FCSP strategies. Thus, we propose the below proposition: 
\begin{proposition}
\normalfont
Given any feasible planning strategy $(\bar{\boldsymbol{u}},\bar{\boldsymbol{w}})$, if we solve \eqref{Reformulated Problem} by fixing $\boldsymbol{u}$ and $\boldsymbol{w}$ with $\bar{\boldsymbol{u}}$ and $\bar{\boldsymbol{w}}$, then the dual variable of the constraint \eqref{ref-dual constraint} regarding scenario $s$ under Period $\gamma$, denoted as $\delta^s_\gamma$, characterizes the worst-case probability for $\boldsymbol{\theta}^{s}_\gamma$ within the DDAS $\mathcal{A}^\theta_\gamma(\bar{\boldsymbol{u}}^{\rm ch})$. Mathematically, we have:
\end{proposition}
\vspace{-3ex}
\begin{align}
    \mathbb{P}_\gamma^{\rm worst}(\tilde{\boldsymbol{\theta}}_{\gamma}\!=\!\boldsymbol{\theta}_{\gamma}^s|\bar{\boldsymbol{u}},\bar{\boldsymbol{w}})\!= \!{\delta}_{\gamma}^{s} 
    \quad \forall s\!\in\!\mathcal{S},\gamma\in\Gamma
\end{align}
\vspace{-2ex}

As constraints \eqref{ref-dual constraint} in the reformulated model is the dual for the DDASs in the original problem, this proposition naturally holds.  In the next section, the reliability of counterpart strategies will be tested based on the derived worst-case distributions.

\subsubsection{Value of Decision-Dependent Distributionally Robust Solution (VD$^3$RS)}
To quantitatively exhibit the benefits of incorporating decision-dependent charging demands into the DRO framework,
a metric called VD$^3$RS is defined. This metric is adapted from the value of decision-dependent stochastic programming solutions employed in the context of SO with DDU 
\cite{zhan2016generation}. 
Given the planning strategy $(\boldsymbol{{u}}^{\rm DIU},
\boldsymbol{{w}}^{\rm DIU})$ derived from the traditional DRO-based FCSP model where DDU of EV adoption is not accounted, VD$^3$RS is mathematically defined as follows:

\vspace{-2.5ex}
\begin{small}
\begin{align}
    \text{V$^3$DRS} = 
\frac{F^{\rm DDU}(\boldsymbol{{u}}^{\rm DIU},
\boldsymbol{{w}}^{\rm DIU})-F^{\rm DDU}(\boldsymbol{{u}^*},\boldsymbol{{w}^*})}
{F^{\rm DDU}(\boldsymbol{{u}^*},\boldsymbol{{w}^*})}
\ge 0
\end{align}
\end{small}
where $(\boldsymbol{{u}^*},\boldsymbol{{w}^*})$ is the optimal planning strategy for the D$^3$R-FCSP model \eqref{compact-instage}. 
The physical implication of the metric is the greatest expected value gained by incorporating the DDU into the DRO framework. In order to ensure a fair comparison across diverse parameters, we normalize the metric through dividing it by the optimal value of D$^3$R-FCSP. 
In Section V, sensitivity analyses will be conducted based on V$^3$DRS to intuitively showcase the effectiveness of our method in various traffic contexts. 

\section{Case Studies}
\subsection{Test Systems and Parameter Setting}




This section presents numerical tests of the proposed method. The Sioux Falls Network \cite{zhou2020robust} is considered to simulate a highway TN, which is coupled with a 110 kV DN \cite{zhang2017second}, as shown in Fig. \ref{Coupled TDN}. The Sioux Falls network has 24 nodes and 76 directed arcs, and the OD pair information and base traffic flow can be found in \cite{zhou2020robust}.  
The TN is divided into three districts, District A, B and C, and the original EV adoption rates are all set at 10\%. The three districts have different EV adoption potential and sensitivity toward FCS installation. Specifically, District A, B and C have high, median and low basic diffusion rates $a_{od,\gamma}$ of 0.04, 0.02 and 0.01 with IF $\Delta d_{od,\gamma}$ of 0.04, 0.02 and 0, respectively.  
EVs in the TDN have identical driving ranges of 360 miles, with an energy consumption of 0.24 kWh/mi. The lower threshold for EV drivers to recharge their EVs is set at 20\%, and thus there are 42 OD pairs left for the study after eliminating the pairs that are too short. For CSs, the rated charging power of each CS is 120 kW with a 92\% charging efficiency. The maximum number of CSs in a single FCS is 90. 
The 110 kV DN has 14 nodes and 13 distribution lines, and the lower and upper limits of node voltage are set at 0.95 and 1.05 p.u., respectively. Other network parameters can be found in \cite{zhang2017second}. TN nodes that do not directly connect to the DN are assumed to be connected to the geographically nearest node in the DN, with a distance of 10\% of the distance to the nearest DN bus \cite{zhang2017second}. The planning horizon is set as 6 years, with three planning periods of 2 years each.

For strategic uncertainty defined by DDASs, we assume that 
robustness parameters for the first- and second moment of EV adoption rates, namely, $\varepsilon_{od,\gamma}^\mu, \check\varepsilon_{od,\gamma}^\upsilon$ and $\hat\varepsilon_{od,\gamma}^\upsilon$, are set at 15\% of the expected EV adoption rate, 0.85 and 1.15, respectively.
For each OD path in each period, 20 scenarios are generated for the supporting set of EV adoption rates.
Regarding operational uncertainties, 8 representative days with 1-hour resolution are generated to simulate the seasonal and weekly patterns of traffic flows, PV output and conventional loads  \cite{zhang2017second},  \cite{wang2016fractal}.

\begin{figure}
    \centering
    \includegraphics[width=9cm,height=5.3cm]{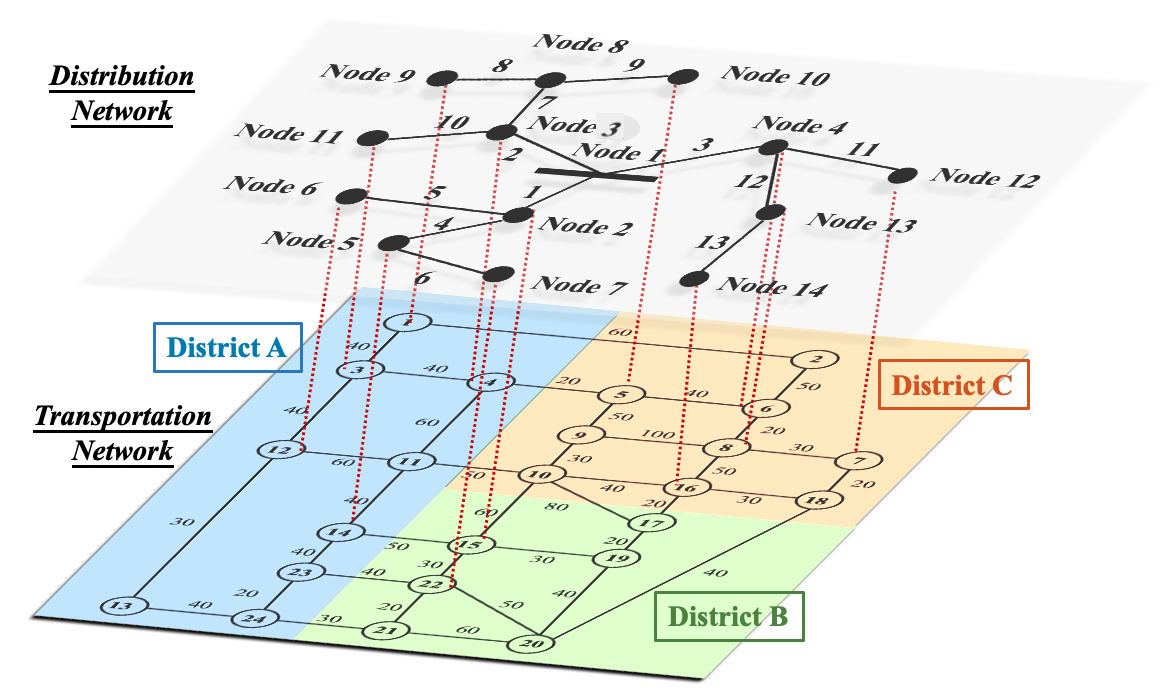}
    \caption{Topology of the test TDN and the coupling relationship}
    \label{Coupled TDN}
\end{figure}


The investment cost for each FCS at a new location is \$1630,000, while each CS with a 120 kW capacity will cost \$100,000. The investment costs for on-site PVs and ESSs are \$883/kW and \$300/kW, and their maximum installed capacities are 6MW and 3MWh, respectively. The charging and discharging power capacities for ESSs is 0.4 and 0.5 of their installed capacities. For DN assets, the per-unit length cost of a distribution line is set at \$120/kVA$\cdot$km, and the cost of substation expansion is set at \$788/kVA. The surplus substation capacity for FCSs at TN nodes is 10MVA. The electricity purchase cost from the main grid is \$0.094/kWh. Penalty costs for unsatisfied EV charging demand, PV curtailment, and load shedding are set at 100\$/kWh, 30\$/kWh, and 50\$/kWh, respectively. The interest rate is 0.06. 

The computation is implemented in Julia using JuMP \cite{dunning2017jump} and solved with Gurobi 11.5 on an Intel Core i5-6500 CPU with 16 GB RAM PC. The convergence tolerance is set at 0.01\%.

\begin{table}[]
\centering
\caption{Different prior assumptions of five strategies}
\label{5-Strategies}
\begin{tabular}{lcc|cc}
\hline\hline
                  & \multicolumn{2}{c}{\textbf{Uncertainty Modeling}} & \multicolumn{2}{c}{\textbf{On-Site Resources}} \\ \cline{2-5} 
\textbf{Strategy} & \textbf{DDU}              & \textbf{Ambiguity}             & \textbf{PV}                & \textbf{ESS}               \\ \hline
I (D$^3$R-FCSP)        & \Checkmark                & \Checkmark                     & \Checkmark                 & \Checkmark                 \\
II (DR-FCSP)      & \XSolid                & \Checkmark                     & \Checkmark                 & \Checkmark                 \\
III (SDD-FCSP)      & \XSolid                & \Checkmark                     & \Checkmark                 & \Checkmark                 \\ 
IV       & \Checkmark                & \Checkmark                     & \Checkmark                 & \XSolid                 \\
V        & \Checkmark                & \Checkmark                     & \XSolid                 & \XSolid                 \\ \hline\hline
\end{tabular}
\vspace{-2.5ex}
\end{table}

\begin{table*}[]
\centering
\caption{Planning Results of Counterpart Strategies}
\label{In-sample performance}
\setlength{\tabcolsep}{1.3mm}{
\begin{tabular}{c|ccc|ccc|ccccc|cc|c}
\hline \hline
                  & \multicolumn{3}{c|}{\textbf{No. of New FCSs/CSs}} & \multicolumn{3}{c|}{\textbf{Expanded Lines}}                                                                                     & \multicolumn{5}{c|}{\textbf{Investment Cost (\$10$^7$)}} & \multicolumn{2}{c|}{\textbf{Operation Cost (\$10$^7$)}} & \textbf{Total Cost}                \\ \cline{2-14}
\textbf{Strategy} & Period 1        & Period 2              & Period 3              & Period 1 & Period 2                                                        & Period 3                                                          & FCS         & PV          & ESS        & Substation       & Line        & Electricity                          & Penalty                         & \textbf{(\$10$^7$)} \\ \hline
I                 & 10 / 366        & 2 / 184        & 3 / 259        & -        & \#2                                                      & \#7                                                        & 6.15        & 2.26        & 0.82       & 0.26             & 11.75       & 57.10                                & 0                               & 78.33                              \\
II                & 8 / 357         & 1 /150         & 3 / 225        & -        & -                                                        & \#2                                                        & 5.88        & 2.08        & 0.63       & 0.32             & 6.06        & 53.02                                & 0                               & 67.99                              \\
III               & 10 / 363        & 2 / 162        & 2 / 226        & -        & -                                                        & \#2                                                        & 5.92        & 2.24        & 0.87       & 0.23             & 5.29        & 55.80                                & 0.16                            & 70.21                              \\
IV                & 10 / 374        & 5 / 204        & 8 / 271        & \#2      & \#3, \#7                                                 & \#12                                                       & 7.02        & 2.92        & -          & 3.35             & 26.81       & 63.31                                & 3.34                            & 106.74                             \\
V                 & 11 / 389        & 9 / 261        & 4 / 207        & \#2, \#7 & \#3, \#5 & \#4, \#9, \#12 & 9.75        & -           & -          & 4.33             & 35.20       & 74.48                                & 9.83                            & 133.59                             \\ \hline \hline
\end{tabular}}
\vspace{-2ex}
\end{table*}

\vspace{-1.5ex}
\subsection{Planning Results and In-Sample Performances}

To demonstrate the significance of accurately modeling the DDU and the ambiguity of progressive EV diffusion, three FCSP strategies with distinct prior assumptions are compared. As the first three rows of TABLE \ref{5-Strategies} show, Strategy I is derived from our proposed D$^3$R-FCSP model, while Strategy II is derived from the distributionally robust FCSP (DR-FCSP) model without accounting for the interdependence between FCS allocation and EV adoption. 
Strategy III is derived from the stochastic decision-dependent FCSP (SDD-FCSP) model with empirical PDFs. 
The FCS deployments and planning results are presented in Fig. \ref{allocation results} and TABLE \ref{In-sample performance}, respectively. 

First, we analyze the in-sample performances of three strategies based on their respective prior assumptions. 
As shown in Fig. \ref{allocation results}, 
despite the similar FCS deployments at the end of Period 3, 
there are significant variations in the timing of FCS allocation and installed capacities. 
For instance, Strategy I installs the highest number of FCSs/CSs, and tends to allocate more FCSs in earlier periods. 
In contrast, Strategy II has installed the fewest FCSs/CSs, as its ignorance toward the FCSs' indirect network effects leads to a comparatively low expectation of EV adoption. 
Due to the optimism toward the empirical PDFs, Strategy III exhibits slightly delayed FCS allocations in certain nodes and fewer CSs compared to Strategy I. 
In addition, the planning results in TABLE \ref{In-sample performance} demonstrate that Strategy I also has higher investments in on-site PV and ESSs, and additionally expands Line 7 to alleviate grid congestion. 
Notably, while Strategy II and III appear to have lower investment and total costs, their security issues and actual cost-efficiency cannot be fairly reflected due to inadequate considerations, as will be discussed in the following subsections. 

\begin{figure}
    \centering
    \includegraphics[width=8.3cm,height=5.4cm]{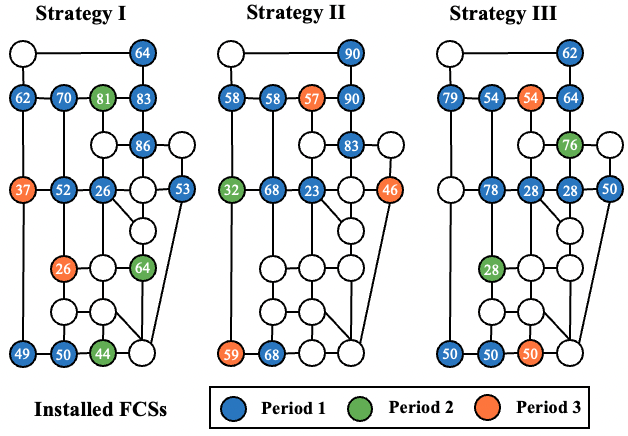}
    \caption{FCS deployment results under Strategy I, II, and III, where the numbers in the circles are the installed CS number at the end of Period 3.}
    \label{allocation results}
    \vspace{-2ex}
\end{figure}

\subsection{Out-of-Sample Performances: Expected EV Diffusion and Potential Security Issues}
To facilitate a fair comparison between the three strategies and reveal their real-world risks, out-of-sample tests are conducted. 
Three test sets are generated by simulating different EV traffic patterns that might occur in the real world. 
The scenarios in the first set are generated from the worst-case PDFs within DDASs (derived by Proposition 2) and referred to as WCD. Another set uses random PDFs within the DDASs to generate scenarios, labeled as RGD, while the third set adopts the empirical PDF for scenario generation and is labeled as ED. Additionally, Strategies IV and V are generated by not permitting the installation of on-site PV and/or ESSs. Due to the page limit, the analyses regarding the contribution of on-site resources will be presented and analyzed in greater detail in Appendix \ref{onsite}.

\begin{table}[]
\centering
\caption{Expected Covered EVs ($\gamma=3,d=4,t=7$) and Load Shedding ($\gamma=3,d=4$), where TF:  \normalfont{traffic flow (\%)} and IF: \normalfont{incentive factor (p.u.)} }
\label{OOSP}
\setlength{\tabcolsep}{1.8mm}{
\begin{tabular}{ccccc|ccc}
\hline\hline
\multicolumn{2}{l}{\textbf{}}       & \multicolumn{3}{c|}{\textbf{Covered EVs (\%)}} & \multicolumn{3}{c}{\textbf{Load Shedding (kWh)}} \\ \cline{3-8} 
                & \textbf{Strategy} & \textbf{WCD}   & \textbf{RGD}   & \textbf{ED}  & \textbf{WCD}    & \textbf{RGD}   & \textbf{ED}   \\ \hline
\textbf{Case 1} & I                 & 99.8           & 100            & 100          & 0               & 0              & 0             \\
TF=100          & II                & 87.1           & 88.7           & 92.4         & 1251.2          & 1103.6         & 822.9         \\
IF=1            & III               & 93.6           & 95.8           & 98.3          & 186.9           & 83.2           & 0             \\ \hline
\textbf{Case 2} & I'                & 93.9           & 95.3           & 96.2         & 491.8           & 326.2          & 260.6         \\
TF=200          & II'               & 77.5           & 81.6           & 83.8         & 6156.9          & 4179.7         & 1791.0        \\
IF=1            & III'              & 89.1           & 90.9           & 95.3         & 764.5           & 791.1          & 641.4         \\ \hline
\textbf{Case 3} & I''               & 98.9           & 99.6           & 100          & 0               & 0              & 0             \\
TF=100          & II''              & 83.8           & 85.9           & 88.0         & 2079.4          & 1623.2         & 1238.8        \\
IF=2            & III''             & 91.3           & 94.7           & 97.6         & 275.4           & 64.6           & 0             \\ \hline\hline
\end{tabular}}
\vspace{-2ex}
\end{table}

\subsubsection{Covered EVs and Load Shedding}
Table \ref{OOSP} presents the expected EV coverage and daily load shedding for the three test sets during the selected time slot and representative day. 
The base case is labeled as Case 1. Furthermore, to investigate the influences of traffic flow (TF) level and IF on security, two additional cases are conducted, by doubling TF and IF in Case 2 and 3, respectively. 

In Case 1, the proposed Strategy I maintains the highest EV coverage rate and zero load shedding across all test sets. 
In contrast, the EV coverage rates for Strategies II and III decrease by 12.7\% and 6.2\% under WCD and 11.3\% and 5.2\% under RGD, respectively, indicating suboptimal investment. 
Particularly, the load shedding in Strategy II and III indicates line congestion caused by excessive EV charging. 
In Case 2 with doubled TF, unsatisfied EVs and load shedding become severe under both Strategy II and III, highlighting the higher risks of neglecting DDU and the ambiguity of EV diffusion under heavy traffic. 
Notably, load shedding occurs across all strategies in Case 2, indicating that the traffic level has surpassed the maximum capability the network can accommodate. 
In Case 3 with doubled IF, Strategy I can still achieve high EV coverage and zero load shedding.  
In contrast, despite resorting to intense load shedding, 
Strategy II and III still exhibit deteriorated EV coverage. 
These results demonstrate the superiority of our proposed model in terms of reliability and security, especially in systems with heavier traffic and heightened sensitivity to charging opportunities.   



\subsubsection{Voltage Profiles and Line Ratings}
The operational security of the coupled DN is also crucial in evaluating the effectiveness of FCSP strategies.
To more intuitively show the charging pressures exerted on the DN, we relax the limits on nodal voltage and line ratings and then re-run the second-stage operation model under WCD set with the first-stage planning decisions held fixed. 
Specifically, we select time slot $t=20$ when there is no solar radiation, so FCSs become more dependent on on-site ESSs and the grid. 
Fig. \ref{Voltage Profile} presents the nodal voltage profiles for three planning periods. Fig. \ref{Line Rating} shows the line ratings for Period 3, where solid lines represent the expected values and boxplots depict the spans of the line ratings under the WCD set.

While Strategy I can ensure all nodal voltages and line ratings are within the limits with sufficient security margins, 
Strategy II and III both exhibit security issues. 
In Strategy II, the significant voltage dip at DN Node 14 and the high risk of overloading in Line 3 are attributed to the deployment of high-capacity FCSs in the downstream area of Node 4. Unlike Strategy I, which distributes more FCSs with lower capacity to disperse the charging pressure so as to defer the expansion of Line 3, Strategy II concentrates the high-capacity FCSs at TN nodes 2, 6, and 8 (as shown in Fig. \ref{allocation results}).
Additionally, the delayed expansion of Line 7 in Strategy II results in severe overloading issues due to the unexpectedly excessive charging demands. 
Similarly, Strategy III also exhibits a risk of overloading Line 7 due to the delayed expansion. This potentially results from its overoptimism toward the empirical PDFs and the underestimation of some high EV adoption scenarios.  

\subsubsection{EV Diffusion Process}
Finally, the projected EV diffusion processes are compared in TABLE \ref{EV diffusion}. Three OD pairs are selected from each district.
For OD-pair 3-19 in District A, Strategy I deploys 4 FCSs along the path, leading to an initial boost in the EV adoption rate during Period 1. This increased EV adoption rate then reinforces the decision-makers' inclination to deploy more FCSs along the same path to disperse the charging pressure. Consequently, there is a higher projected EV adoption than those in Strategy II and III, demonstrating a positive feedback loop. 
Therefore, in the case of Strategy II, although the initial deployment of FCSs is identical to that of Strategy I, the delays and reductions in FCS deployment result in a slower rate of EV diffusion. And this can also explain the security issues discussed before.
On the other hand, Strategy III's optimism about the empirical distribution of EV adoption leads to fewer FCS installations from the start. This, in turn, also contributes to a slower rate of EV diffusion. 
Similar trends can be observed for OD pair 20-3 in District B, albeit with smaller magnitudes. 
Another notable observation from the table is that the timing of FCS deployment has a significant impact on the EV diffusion process. 
The results for OD pairs 3-19 and 20-3 indicate that early FCS deployment has a cumulative effect and contributes more to the overall adoption of EVs.  


Thus, in addition to improving EV coverage and deferring costly grid expansion without deteriorating DN security, our proposed method can also speed up the diffusion of EVs by enhancing the positive feedback loop.
Conversely, failing to consider the indirect network effects of FCS installation and ambiguous PDFs can pose challenges in meeting unexpectedly excessive charging demands and compromise the security of the operation. 
These results not only highlight the robustness and reliability of our proposed model, but also underscore the significance of an informed and adaptive FCSP strategy for better promoting EV diffusion.



\begin{figure}
    \centering
    \includegraphics[width=8cm,height=4cm]{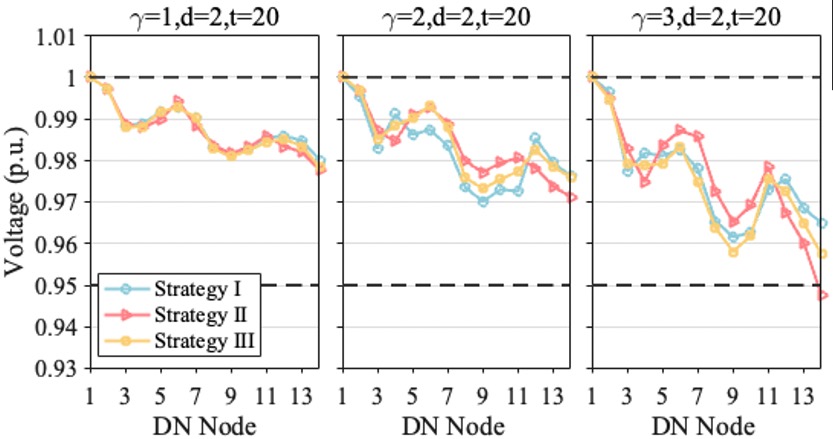}
    \vspace{-2ex}
    \caption{Voltage profiles at three planning periods}
    \vspace{-2ex}
    \label{Voltage Profile}
\end{figure}

\begin{figure}
    \centering
    \includegraphics[width=8cm,height=4cm]{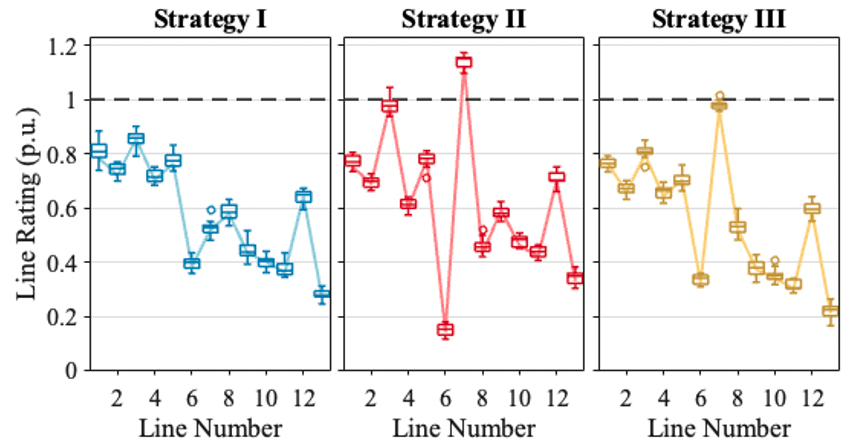}
    \vspace{-2ex}
    \caption{Line ratings of the DN at $\gamma=3, d=3,t=20$}
    \vspace{-3ex}
    \label{Line Rating}
\end{figure}

\subsection{Sensitivity Analysis: The Computation of VD$^3$RS}
To comprehensively evaluate the effectiveness of our strategy across various external conditions in terms of monetary value, the VD$^3$RS introduced in Section IV.C is computed. We perturb the values of IF and TF and present the results in Fig. \ref{VD3RS}, which offer the following insights and implications.

Given a fixed TF, the metric grows monotonically with the increasing IF. 
This result is trivial and intuitive, as the stronger sensitivity the drivers have to the FCSs' presence, the higher regrets the decision makers would have for not considering the incentivized charging demands when deriving investment strategy. Consequently, heightened penalties in the operation stage will lead to higher VD$^3$RS. 
Specifically, when the TF is set at 125\%, the incorporation of DDU yields expected cost savings of 25.9\% and 30.3\% under IF = 1 and 2 p.u., respectively.

\begin{table}[]
\centering
\caption{Expected EV Diffusion of Counterpart Strategies}
\label{EV diffusion}
\setlength{\tabcolsep}{0.3mm}{
\begin{tabular}{cc|ccc|ccc}
\hline\hline
\textbf{OD Pair} & \textbf{Strategy} & \multicolumn{3}{c|}{\textbf{FCS Location}} & \multicolumn{3}{c}{\textbf{EV Adoption (\%)}} \\ \cline{3-8} 
(Shortest Path)  &                   & Period 1     & Period 2     & Period 3     & Period 1         & Period 2         & Period 3         \\ \hline
\textbf{3-19}    & I                 & 3,4,6,8      & 5,19         & -            & 16.48            & 23.83            & 30.53            \\
(3-4-5-6-        & II                & 3,4.6,8      & -            & 5            & 16.48            & 22.49            & 28.69            \\
8-16-17-19)      & III               & 3,4,6        & 8            & 5            & 15.76            & 21.83            & 28.08            \\ \hline
\textbf{20-3}    & I                 & 3,13,24      & 21           & 12           & 12.34            & 14.79            & 17.35            \\
(20-21-24-       & II                & 3,24         & 12           & 13           & 12.16            & 14.44            & 16.84            \\
13-12-3)         & III               & 3,13,24      & -            & 21           & 12.34            & 14.62            & 17.01            \\ \hline
\textbf{2-15}    & I                 & 2,6,8        & 19           & -            & 11.50            & 13.00            & 14.50            \\
(2-6-8-16-       & II                & 2.6,8        & -            & -            & 11.50            & 13.00            & 14.50            \\
17-19-15)        & III               & 2,6,16       & 8            & -            & 11.50            & 13.00            & 14.50            \\ \hline\hline
\end{tabular}}
\end{table}


On the other hand, when increasing the TF under a fixed IF, the VD$^3$RS grows at first and reaches a peak around TF = 150\%. 
However, it is surprising that as the TF increases further, the VD$^3$RS metric exhibits a declining trend. 
Referring to TABLE \ref{OOSP}, the coexistence of load shedding and uncovered EVs in both Strategy I and II under Case 2 indicates that, under heavy traffic conditions, power supply capacity from the grid side becomes the predominant limiting factor. 
Therefore, even with our more informed FCSP strategy, it is difficult to offset the penalties resulting from uncovered EVs and security violations.

\begin{figure}
    \centering
    \includegraphics[width=8.5cm,height=3.9cm]{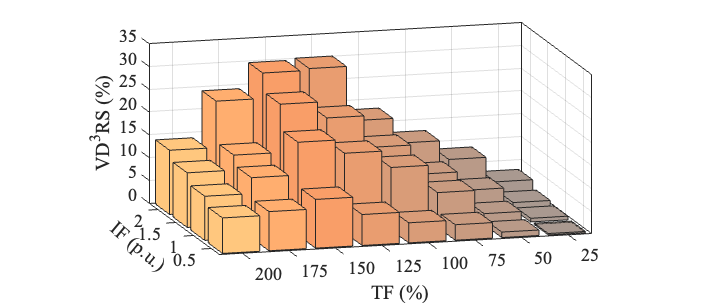}
    \caption{VD$^3$RS under different incentive factors and traffic flow levels}
    \label{VD3RS}
    \vspace{-2ex}
\end{figure}

In sum, the computation of VD$^3$RS verifies the superiority of the D$^3$R-FCSP model in most cases from a monetary standpoint, as our method can effectively offset the risk of high operation costs, despite the seemingly higher first-stage investment costs. 
In practice, considering different TDNs have varying traffic patterns and drivers have diverse sensitivities to FCS installation, VD$^3$RS can offer valuable insights for decision makers regarding the value of a more informed EV diffusion model under various conditions. 

\section{Conclusion}


Recognizing that the interplay between EV diffusion and FCS deployment is a classic ``chicken-and-egg" problem, this paper introduces a novel two-stage D$^3$R-FCSP model to meet the decision-dependent charging demands through adaptive investment. 
The proposed model integrates DDASs to effectively address the key characteristics of EV diffusion: 1) the DDU of EV adoption that can be endogenously reshaped by FCS locations, and 2) the ambiguous PDFs resulting from the lack of information. 
Additionally, the MCACPC model is adopted to ensure the feasibility of locations and capacity of FCSs by capturing EV recharging patterns. 
To manage the computational complexity, we present an equivalent single-level MILP reformulation of the D$^3$R-FCSP model. 
In numerical studies, extensive out-of-sample evaluations demonstrate that our method can better cover the reshaped EV charging demands while maintaining sufficient security margins, by deploying RES-powered FCSs in an adaptive and informed manner. 
Moreover, the computation of VD$^3$RS  metric offers monetary insights of the explicit quantification of DDUs under various traffic conditions. 
Notably, results also exhibit an accelerated EV diffusion process under our strategy, through 
enhancing the mutual reinforcement between EV adoption and FCS construction. 
These findings can potentially contribute to a more reliable and faster transition to transportation electrification. 

{\footnotesize
\bibliographystyle{IEEEtran}
\bibliography{reference}}

\appendices
\setcounter{table}{0}
\setcounter{figure}{0}
\setcounter{equation}{0}
\renewcommand\theequation{A-\arabic{equation}}

\section{Procedure for generating candidate FCS locations}
\label{ACPC-algo}

Considering limited range of EVs, the fundamental logic of the MCACPC 
is that all arcs on all paths should be traversable by the EVs without depleting their battery on the road. 
Specifically, let $\mathcal{A}_{od}^{\rm TN}$ be the set of ordered TN nodes on the shortest path for OD pair $od$. For each arc $(j,k)$ of $od$, we define the set $\mathcal{K}_{od}(j,k)$ containing all candidate FCS nodes which allowing an EV to traverse the arc $(j,k)$ without depleting its battery before reaching node $k$. 
In Fig. \ref{Toy}, assuming the EV driving range is $D^{\rm ev} = 100$ mile, a driver traveling along the OD pair $(1,4)$ who wants to traverse  the arc $(3,4)$ without running out of battery must recharge the battery either at node $2$ or $3$. Therefore, we have $\mathcal{K}_{od}(3,4) = \{2,3\}$. Similarly, to cover the backward path $(3,2)$, the driver must recharge at node $4$ or $3$, so $\mathcal{K}_{od}(3,2) = \{3,4\}$. 
This process is repeated for each arc on each path. 
Algorithm \ref{algorithm 1}  presents the pseudocode for generating the candidate set $\mathcal{K}_{od}(j,k)$, where $\text{dis}_{od}(j,k)$ denotes the length of arc $(j,k)$. 

\begin{algorithm}
\caption{pseudocode for Generating $\mathcal{K}_{od}(j,k)$ }
\label{algorithm 1}
\For{$od=1$ \KwTo $|\mathcal{Q}|$}{
\For{
$\forall(j,k)\in\mathcal{A}_{od}^{\rm tr} \wedge j\le k$
}
{
\uIf {$i \le j \ \wedge \ {\rm dis}_{od}(i,k)\le L$} {Add $i$ into $\mathcal{K}_{od}(j,k)$
} 
\ElseIf {$i\ge k\ \wedge\ {\rm dis}_{od}(i,r_{od})\!+\!{\rm dis}_{od}(r_{od},k)\!\!\le\!\! L$} 
{
Add $i$ into $\mathcal{K}_{od}(j,k)$
}
}
\For{ $\forall(k,j)\in\mathcal{A}_{od}^{\rm tr} \wedge j\le k$ }{

\uIf {$i\ge k \wedge {\rm dis}_{od}(i,k)\le L$} {Add $i$ into $\mathcal{K}_{od}(k,j)$
    
} \ElseIf {$i\ge j\ \wedge\ {\rm dis}_{od}(i,s_{od})\!+\!{\rm dis}_{od}(s_{od},j)\!\!\le\!\! L$} {
    Add $i$ into $\mathcal{K}_{od}(k,j)$
}
}
}
\end{algorithm}

\section{Proof of Proposition 1}

\begin{proof}
First, we rewrite the $\gamma$-th inner superior operation with regard to $\mathbb{P}_\gamma$ as follows:

\vspace{-2ex}
\begin{small}
\begin{subequations}
\begin{align}
    {\mathop{\rm sup}}\ 
    &{\rm E}_{\mathbb{P}_\gamma}[\varphi(\boldsymbol{x}_\gamma,\boldsymbol{z}_\gamma,\boldsymbol{\tilde{\theta}}_\gamma)]
    =
    {\mathop{\rm max}_{\pi_{\gamma}^{s} }}
    \sum_{s \in \mathcal{S}_\gamma}\!\pi_{\gamma}^{s}\varphi_{\gamma}(\boldsymbol{x}_\gamma,\boldsymbol{z}_\gamma,\boldsymbol{\theta}_\gamma^s)
    \label{ref-obj1-Ap}
    \\
    &{\rm s.t.} \qquad \qquad \qquad
    \sum_{s\in \mathcal{S}_\gamma} \pi_\gamma^{s}=\boldsymbol{1}
    \qquad 
    (\kappa_\gamma)
    \\
    - \!\! &\!\sum_{s\in\mathcal{S}_\gamma}\!
    \pi_\gamma^s\theta_{od,\gamma}^s
    \!\!\le\!\!
    -
    (\bar{\mu}_{od,\gamma}\!-\!\varepsilon^{\mu}_{od,\gamma}) \!-\!
    \sum_{r=1}^{\gamma-1}
    \sum_{i\in\mathcal{N}_{od}^{\rm TN}} \!\!
    \Delta \mu_{od,i,\gamma} x^{\rm ch}_{i,r}
    \quad
    \notag
    \\
&\qquad\qquad\qquad\qquad\qquad\qquad(\alpha^\mu_{od,\gamma})\qquad \forall od \in \mathcal{Q}
    \\
    & \sum_{s\in\mathcal{S}_\gamma}
    \!\!\pi_\gamma^{s}\theta_{od,\gamma}^s 
    \!\le\! 
    (\bar{\mu}_{od,\gamma}\! +\!\varepsilon^{\mu}_{od,\gamma}) \!+\!
    \sum_{r=1}^{\gamma-1}
    \sum_{i\in\mathcal{N}_{od}^{\rm TN}} \!\!\Delta \mu_{od,i,\gamma} x^{\rm ch}_{i,r}
    \quad 
    \notag
    \\
&\qquad\qquad\qquad\qquad\qquad\qquad
(\beta^\mu_{od,\gamma})
\qquad \forall od \in \mathcal{Q}
    \\
-\!\!\! &\sum_{s\in\mathcal{S}_\gamma}\!\!\pi_\gamma^{s}(\theta_{od,\gamma}^s)^2
    \!\le\!\!
    -(\bar{\mu}_{od,\gamma}^2 \!\!+\!\bar{\sigma}_{od,\gamma}^2) 
    (1\!+\! \!\!\!\! \!\sum_{i\in\mathcal{N}^{\rm TN}_{od}}
    \!\!\!\!\!
    \Delta \upsilon_{od,i,\gamma}
    x^{\rm ch}_{i,\gamma-1})\check{\varepsilon}_{od,\gamma}^{\upsilon}
    \notag
    \\
&\qquad\qquad\qquad\qquad\qquad\qquad
(\alpha_{od,\gamma}^\upsilon)
\qquad \forall od \in \mathcal{Q}
    \\
&\sum_{s\in\mathcal{S}_\gamma}\!\!\pi_\gamma^{s}(\theta_{od,\gamma}^s)^2
    \!\le\! \!
    (\bar{\mu}_{od,\gamma}^2 \!\!+\!\bar{\sigma}_{od,\gamma}^2) 
    (1\!+\! \!\!\!\sum_{i\in\mathcal{N}^{\rm TN}_{od}}
    \!\!\!\!\!
    \Delta \upsilon_{od,i,\gamma}
    x^{\rm ch}_{i,\gamma-1})\hat{\varepsilon}_{od,\gamma}^{\upsilon}
    \notag
    \\
&\qquad\qquad\qquad\qquad\qquad\qquad
(\beta_{od,\gamma}^\upsilon)
\qquad \forall od \in \mathcal{Q}
\end{align}
\label{reformulation1}
\end{subequations}
\end{small}
\noindent  where the variable in the bracket after each constraint corresponds to the dual variable of the constraint. 
Since by assumption there always exists a relative interior within the feasible set of \eqref{reformulation1} given any feasible first-stage solution $\boldsymbol{x}^{\rm ch}_{i,\gamma}$, the Slater's condition holds. Furthermore, as the original model \eqref{compact-instage} has complete recourse, we can use the strong duality to reformulate \eqref{reformulation1} to its dual problem equivalently:

\vspace{-1ex}
\begin{small}
\begin{subequations}
\begin{align}
     &\min_{
    {\kappa}_\gamma,
    \boldsymbol{\alpha}_\gamma\ge 0, 
    \atop
    \boldsymbol{\beta}_\gamma\ge 0
    } 
    \kappa_\gamma
    +\!\!\!
    \sum_{od\in\mathcal{Q}} \!\!
    \big[
    (\varepsilon_{od,\gamma}^{\mu}\!-\!\bar{\mu}_{od,\gamma})\alpha^\mu_{od,\gamma} -
    \notag
    \\
    &
    \sum_{r=1}^{\gamma-1}
    \sum_{i\in\mathcal{N}_{od}^{\rm TN}} \!\!\bar{\mu}_{od,\gamma} \Delta \mu_{od,i,\gamma} \alpha^\mu_{od,\gamma} x^{\rm ch}_{i,r}
    \!-\! (\bar{\mu}_{od,\gamma}^{2}\!+\!\bar{\sigma}_{od,\gamma}^{2}) \check{\varepsilon}_{od,\gamma}^{\upsilon}
    \alpha^\upsilon_{od,\gamma}
    -\!
    \notag
    \\
    & \sum_{i\in\mathcal{N}_{od}^{\rm TN}}\!\!\!\! (\bar{\mu}_{od,\gamma}^{2}\!+\!\bar{\sigma}_{od,\gamma}^{2})\!
    \Delta\upsilon_{od,i,\gamma} \check{\varepsilon}_{od,\gamma}^{\upsilon} \alpha^\upsilon_{od,\gamma} x^{\rm ch}_{ i,\gamma-1} 
    \!+\! (\bar{\mu}_{od,\gamma} \!\!+\!\varepsilon_{od,\gamma}^{\mu}) \beta^{\mu}_{od,\gamma}
    \notag
    \\
    &+\!
    \sum_{r=1}^{\gamma-1}
    \sum_{i\in\mathcal{N}_{od}^{\rm TN}} \!\!\bar{\mu}_{od,\gamma}\Delta \mu_{od,i,\gamma} \beta^\mu_{od,\gamma} x^{\rm ch}_{i,r} \!+\! (\bar{\mu}_{od,\gamma}^{2}\!+\!\bar{\sigma}_{od,\gamma}^{2})
    \hat{\varepsilon}_{od,\gamma}^{\upsilon}\beta^{\upsilon}_{od,\gamma}
    \notag
    \\
    &\! +\! \sum_{i\in\mathcal{N}_{od}^{\rm TN}} \!\!\!(\bar{\mu}_{od,\gamma}^{2}\!+\!\bar{\sigma}_{od,\gamma}^{2})
    \Delta\upsilon_{od,i,\gamma} \hat{\varepsilon}^\upsilon_{od,\gamma} \beta^{\upsilon}_{od,\gamma} x^{\rm ch}_{i,\gamma} \big]
    \label{proof-obj}
    \\
    & {\rm s.t.:}\ 
    \kappa_{\gamma} + \sum_{{od}\in\mathcal{Q}} \!\theta_{od,\gamma}^s (\beta^{\mu}_{od,\gamma}\!-\!\alpha^{\mu}_{od,\gamma})\! + \!\sum_{od\in\mathcal{Q}}\!(\theta_{od,\gamma}^s)^2(\beta^{\upsilon}_{od,\gamma}\!-\!\alpha^{\upsilon}_{od,\gamma})\notag
    \\
    &\qquad\qquad\qquad\qquad
    \ge 
    \varphi_{\gamma}(\boldsymbol{x}, \boldsymbol{z},\boldsymbol{\theta}_\gamma^s) \qquad \forall s\in\mathcal{S}_\gamma 
    \label{ref-dual-constraint}
    \end{align}
{\normalfont where \eqref{ref-dual-constraint} can be further transformed into the following:}
    \begin{align}
     & 
    \kappa_{\gamma} + \sum_{{od}\in\mathcal{Q}} \!\theta_{od,\gamma}^s (\beta^{\mu}_{od,\gamma}\!-\!\alpha^{\mu}_{od,\gamma})\! + \!\sum_{od\in\mathcal{Q}}\!(\theta_{od,\gamma}^s)^2(\beta^{\upsilon}_{od,\gamma}\!-\!\alpha^{\upsilon}_{od,\gamma})
    \notag
    \\
    &\qquad \qquad\qquad\qquad
    \ge 
    \sum_{d\in\mathcal{D}_\gamma} W_d\boldsymbol{c}_3\boldsymbol{y}_d^s
    \qquad 
    \forall  s\in\mathcal{S}_\gamma 
    \\
     &\qquad\qquad \boldsymbol{C}_d \boldsymbol{y}_d^s \le \boldsymbol{l}_d 
    - \boldsymbol{D}_d\boldsymbol{x}_\gamma - \boldsymbol{E}_d\boldsymbol{z}_\gamma - \boldsymbol{F}_d\boldsymbol{{\theta}}^s_\gamma
    \notag
    \\
    &\qquad\qquad\qquad\qquad\qquad\qquad
    \forall  s\in\mathcal{S}_\gamma, d\in\mathcal{D}_\gamma
    \end{align}
\label{Prop1-Ap}
\end{subequations}
\end{small}
To linearize bilinear terms induced by DDU, we adopt McCormick envelope method by introducing auxiliary variables, i.e., $\nu^{\rm I}_{od,i,\gamma}=\alpha^\mu_{od,\gamma} x^{\rm ch}_{i,\gamma-1}$, $\nu^{\rm I}_{od,i,\gamma}=\alpha^\upsilon_{od,\gamma} x^{\rm ch}_{i,\gamma-1}$, $\nu^{\rm I}_{od,i,\gamma}=\beta^\mu_{od,\gamma} x^{\rm ch}_{i,\gamma-1}$ and $\nu^{\rm I}_{od,i,\gamma}=\beta^\upsilon_{od,\gamma} x^{\rm ch}_{i,\gamma-1}$, and insert them into the \eqref{proof-obj}. The corresponding McCormick envelops, i.e., 
 $\boldsymbol{M}^{\rm I}_{od,i,\gamma}$, $\boldsymbol{M}^{\rm II}_{od,i,\gamma}$, $\boldsymbol{M}^{\rm III}_{od,i,\gamma}$, and $\boldsymbol{M}^{\rm IV}_{od,i,\gamma}$, are also added to define the feasible sets, as \eqref{ref-con2} show (details for this technique can be found in \cite{mccormick1976computability}). By combining the linearized formulation with the first stage, we obtain the final single-level formulation \eqref{Reformulated Problem}. This completes the proof.
\end{proof} 

\begin{algorithm}
\caption{pseudocode for Benders Decomposition Algorithm}\label{BD}
{\bf{Step 1.}}\quad {\bf{Initialization.}} Set lower bound $\rm LB\leftarrow -\infty$, upper bound $\rm UB\leftarrow +\infty$, iteration time $l\leftarrow 0$, and optimality gap $\delta=0.01\%$\;
{\bf{Step 2.}}\quad
\While{$|\frac{\rm UB-LB}{\rm UB}|>\delta$}{
    Update $l=l+1$
\\Solve MP \eqref{MP} to obtain the candidate first-stage solutions, $[\boldsymbol{x}_\gamma^{(l)*},\boldsymbol{z}_\gamma^{(l)*},\forall \gamma\in \Gamma]^{\rm T}$, and the optimal value $V_1^{(l)}$
\\
    Update ${\rm LB} \leftarrow V_1^{(l)}$
    \\
    \For{$\gamma=1$ \KwTo$|\Gamma|$}
    {\For{$s=1$ \KwTo$\mathcal{S}_\gamma$}{Solve $(\gamma,s)$-th SP to obtain the optimal dual variable, $\boldsymbol{\tau}_{\gamma}^{s(l)*}$,
    and the optimal value ${V}_2^s
    (\boldsymbol{x}_\gamma^{(l)*}\!, \boldsymbol{z}_\gamma^{(l)*})$
    Generate a new optimality cut \eqref{MP-optimality cut} based on  $\boldsymbol{\tau}_{\gamma}^{s(l)*}$ and ${V}_2^s
    (\boldsymbol{x}_\gamma^{(l)*}\!, \boldsymbol{z}_\gamma^{(l)*})$
    }}
    Compute \eqref{compact-instage} with fixed first-stage decisions $(\boldsymbol{x}_\gamma^{(l)*}\!, \boldsymbol{z}_\gamma^{(l)*})$, and obtain the optimal value $V_0(\boldsymbol{x}_\gamma^{(l)*}\!, \boldsymbol{z}_\gamma^{(l)*})$
    Update ${\rm{UB}} =$ ${\rm{min}}\{{\rm{UB}},V_0(\boldsymbol{x}_\gamma^{(l)*}\!, \boldsymbol{z}_\gamma^{(l)*})\}$
    }
    {\bf{Step 3.\quad Terminate.}} Return \! $\rm UB$ and \!$[\boldsymbol{x}_\gamma^{(l)*}\!\!,\boldsymbol{z}_\gamma^{(l)*}\!\!,\forall \gamma\!\in\!\!\Gamma]$
\end{algorithm}

\section{Benders Decomposition Algorithm}
\label{app-BD}

The main idea of Benders decomposition is to decompose the original problem into a relaxed master problem (MP) and multiple sub-problems (SPs), which are solved iteratively until convergence, thereby reducing computational burdens. The formulations of the investment MP and period-wise scenario-wise operation SPs are as follows.

\subsubsection{Investment Master Problem}
At $L$-th iteration, the investment MP can be defined as the following MILP problem:

\vspace{-1.5ex}
\begin{small}
\begin{subequations}
\begin{align}
    \text{(MP)} &\qquad 
    \text{Objective Function:} \qquad 
    V_1^{(L)} = \eqref{reform-obj}
    \\ 
    &\text{s.t.:} \qquad\qquad
    \eqref{ref-con1},\eqref{ref-con2}
    \\
    \kappa_{\gamma} & + \sum_{{od}\in\mathcal{Q}} \!\theta_{od,\gamma}^s (\beta^{\mu}_{od,\gamma}\!-\!\alpha^{\mu}_{od,\gamma})\! + \!\sum_{od\in\mathcal{Q}}\!(\theta_{od,\gamma}^s)^2(\beta^{\upsilon}_{od,\gamma}\!-\!\alpha^{\upsilon}_{od,\gamma})
    \label{ref-dual constraint-A}
    \notag
    \\
    &\qquad \qquad\qquad\qquad
    \ge 
    \omega^{s}_{\gamma}
    \qquad 
    \forall \gamma \in \Gamma, s\in\mathcal{S}_\gamma 
     \\ 
    \omega^{s}_{\gamma}
    & \ge
      {V}_2^s
     (\boldsymbol{x}_\gamma^{(l)*}\!, \boldsymbol{z}_\gamma^{(l)*})
    \!-\!
    [D_d(\boldsymbol{x}_\gamma\!-\!\boldsymbol{x}_\gamma^{(l)*})^{\rm T} \!+\!
    E_d(\boldsymbol{z}\!-\!\boldsymbol{z}_\gamma^{(l)*})^{\rm T}]{{\boldsymbol{\tau}_{\gamma}^{s}}^{(l)*}}
    \notag
    \\ 
    &
    \qquad\qquad\qquad
    \forall \gamma \in \Gamma, s\in\mathcal{S}_\gamma,
    l=1,..,L-1
    \label{MP-optimality cut}
    \\
    & 
    \qquad \omega^{s}_{\gamma}
     \ge M_0
     \qquad
    \forall \gamma \in \Gamma, s\in\mathcal{S}_\gamma
\end{align}
\label{MP}
\end{subequations}
\end{small}
where constraints \eqref{MP-optimality cut} represent optimality cuts identified in SPs through $L-1$ iterations based on the first-order Taylor-series approximations of second-stage problem \eqref{compact-second-objective}-\eqref{compact-second-constraint} at candidate optimal solutions $[\boldsymbol{x}_\gamma^{(l)*},\boldsymbol{z}_\gamma^{(l)*}]^{\rm T}(l=1,..,L-1)$. $\omega^s_\gamma$ is introduced as an auxiliary variable to offer an under-approximation of the second-stage optimal value under the realization $\boldsymbol{\theta}^s_\gamma$. 
As the MP is a progressively tightened relaxation of the original problem, its optimal value, $V^{(L)}_1$, offers a nondecreasing lower bound (LB) for \eqref{Reformulated Problem} in each iteration. 

\subsubsection{Scenario-Wise Period-Wise Operation Sub-Problems}
Given the trial investment decision $[\boldsymbol{x}_\gamma^{(L)*}\!, \boldsymbol{z}_\gamma^{(L)*},\forall \gamma\in\Gamma]$ derived by solving MP \eqref{MP} at the $L$-th iteration, the operation SPs corresponding to the EV adoption rate realization of $\boldsymbol{\theta}_\gamma^s$ can be formulated as below:

\vspace{-1.5ex}
\begin{small}
\begin{subequations}
\begin{align}
    \text{($(\gamma,s)$-th SP)}\qquad
    {V}^{s}_{2}(\boldsymbol{x}_\gamma^{(L)*}\!, \boldsymbol{z}_\gamma^{(L)*})
     =
    {\mathop{\rm min}} \!
    \sum_{d\in\mathcal{D}_\gamma} 
    W_d
    & \boldsymbol{c}_3
    \boldsymbol{y}_d^s
    \\
    \text{s.t.:}
    \quad 
    \boldsymbol{C}_d\boldsymbol{y}_d^s
    \le \boldsymbol{l}_d 
     - \boldsymbol{D}_d\boldsymbol{x}_\gamma^{(L)*} - \boldsymbol{E}_d\boldsymbol{z}_\gamma^{(L)*} - \boldsymbol{F}_d &\boldsymbol{\theta}_\gamma^s 
    \notag
    \\ 
    (\boldsymbol{\tau}_\gamma^{s})
    \quad
    \forall d  \in & \mathcal{D}_\gamma
    \label{SP-constraint}
\end{align}
\label{SP}
\end{subequations}
\end{small}
By solving each SP, we acquire the optimal solution, ${V}^{s}_{2}(\boldsymbol{x}_\gamma^{(L)*}\!, \boldsymbol{z}_\gamma^{(L)*})$, and the optimal dual variable corresponding to second-stage constraints \eqref{SP-constraint}, $\boldsymbol{\tau}_\gamma^{s(l)*}$. With $\boldsymbol{\tau}_\gamma^{s(l)*}$, a new optimality cut is formed and fed into the MP for executing the next iteration. Meanwhile, since the trial first-stage decision obtained from the MP in the $L$-th iteration is a suboptimal first-stage solution before reaching convergence, when input into the original model \eqref{compact-instage} as fixed values, it can provide a candidate upper bound (UB). 
In this manner, the MP and SPs are formulated as an MILP and LPs, respectively, which can be directly solved using off-the-shelf solvers. We iteratively compute the MP and SPs until the difference between LB and UB is within a predefined tolerance $\delta=0.01\%$. The pseudocode is presented in Algorithm \ref{BD}.

\section{Contributions of on-site ESSs and PVs}
\label{onsite}

\begin{figure}

\centering

\includegraphics[width=9cm,height=4.7cm]{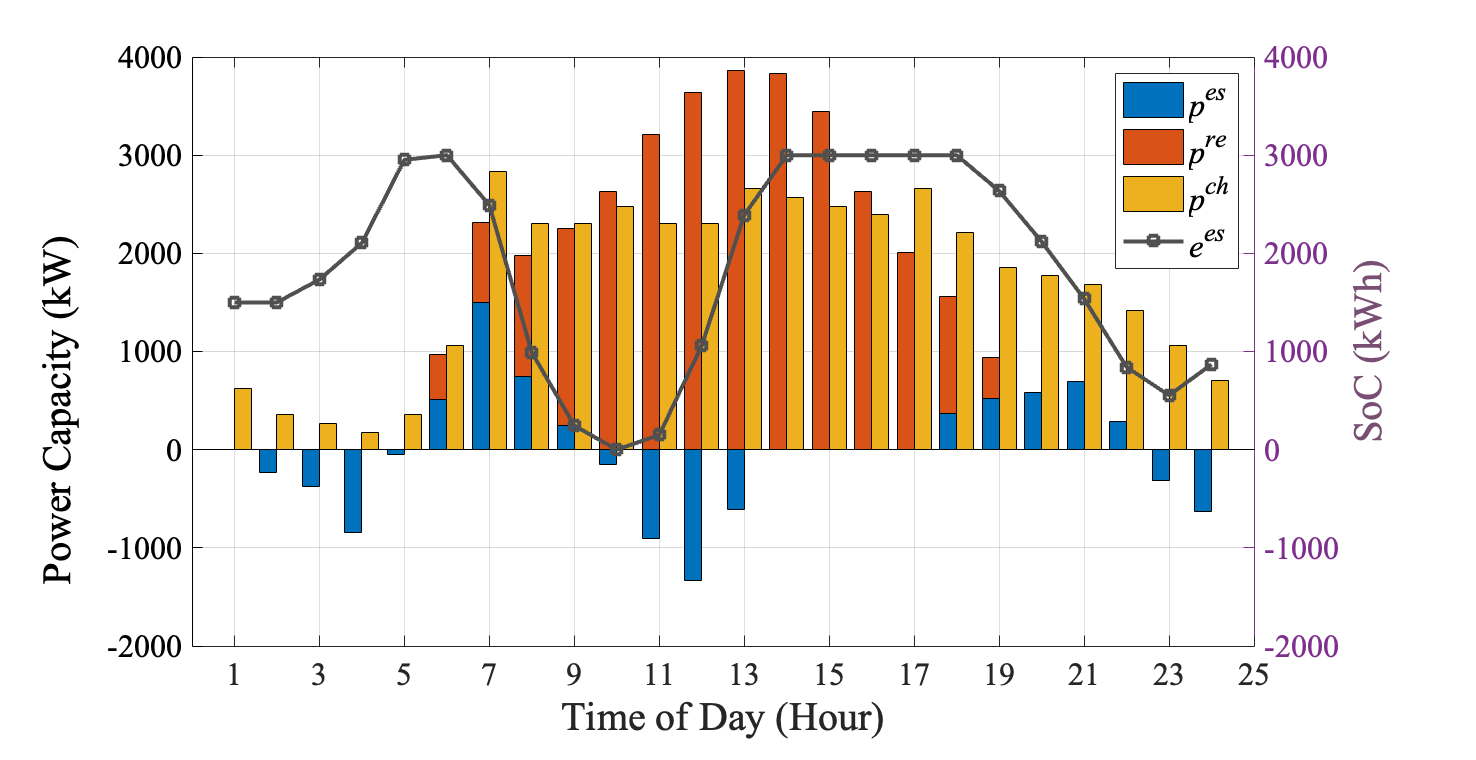}
\begin{center}
{\scriptsize (a) \textbf{Stragy I}: No. of CSs: 32, installed capacity of PV/ESSs: 4 MW/3 MWh }
\end{center}
\includegraphics[width=9cm,height=3.5cm]{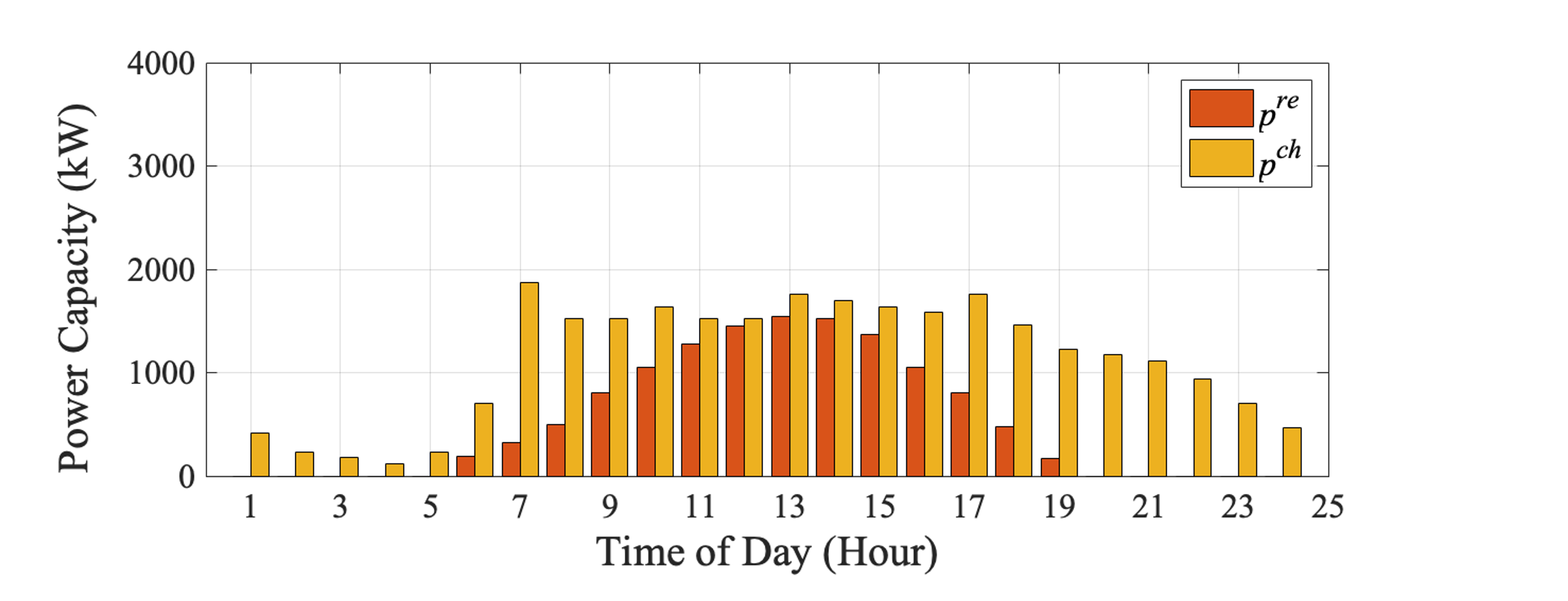}
\begin{center}
    {\scriptsize (b) \textbf{Strategy IV}: No. of CSs: 22, installed capacity of PV: 1.8 MW}
\end{center}
\caption{Daily power and energy flow at TN node 5 ($\gamma=3,d=2$)}
\label{Energy Flow}
\vspace{-3ex}
\end{figure}


To assess the contributions of on-site DERs to EV integration, Strategy IV and V are generated, as defined in TABLE \ref{5-Strategies}. Strategy IV does not consider on-site ESSs, whereas Strategy V exclude both on-site PVs and ESSs. The planning results are presented in the final two rows of TABLE \ref{In-sample performance}.

Compared to Strategy I, Strategies IV and V tend to install more FCSs/CSs in the TN, and conduct more line expansions, showing a more severe line congestion and energy shortage. This results in an increase in both investment and expected operating costs. 
To investigate the rationale behind this, we depict the power and energy flow of the TN node 5 under Strategy I and IV 
in Fig. \ref{Energy Flow}.  
In Fig. \ref{Energy Flow}(a), on-site ESSs in Strategy I effectively stores surplus energy generated during peak sunlight hours (11-13) and light loaded hours (23-5), and then supply power during periods of low sunlight (18-22) and heavily loaded hours (6-9). 
In Fig. \ref{Energy Flow}(b), 
the absence of ESSs in Strategy IV results in fewer installation of PV panels to hedge the risk of curtailment brought on by PV intermittency. 
This, in turn, leads to higher electricity purchase cost from the DN and higher line expansion cost. Furthermore, when both on-site PVs and ESSs are excluded in Strategy V, 7 lines have to be expanded, and FCSs are installed in all TN nodes. Despite the significant investments made, the penalties for load shedding and unmet EV remain substantial. 
This indicates that the TDN's capacity to accommodate EV charging has reached its limit. 

Therefore, by leveraging the economic benefits of on-site PVs and the flexibility of on-site ESSs, the pressure on the DN from EV charging can be significantly reduced. This is achieved by locally meeting EV charging demands, which in turn greatly improves the reliability of the TDN and pushes the boundaries of the TDN's EV integration capacity, without the need for more expensive infrastructure expansions.

\end{document}